\documentclass[british,english,aps, twocolumn,groupedaddress, superscriptaddress, showkeys]{revtex4-1}
\pdfoutput=1

\makeatletter
\renewcommand{\p@subsection}{}
\renewcommand{\p@subsubsection}{}
\makeatother

\usepackage{ae,aecompl}
\usepackage[LGR,T1]{fontenc}
\usepackage[latin9]{inputenc}
\usepackage{float}
\usepackage{textcomp}
\usepackage{siunitx}
\usepackage{amsmath}
\usepackage{amssymb}
\usepackage{graphicx}
\usepackage{subscript}
\usepackage{babel}
\usepackage{hyperref}
\usepackage{gensymb}
\usepackage{mathtools}
\usepackage{subfiles}
\usepackage[table,xcdraw]{xcolor}
\usepackage{tabularx}
\usepackage{longtable}

\usepackage[capitalise]{cleveref} 

\makeatletter

\DeclareRobustCommand{\greektext}{%
  \fontencoding{LGR}\selectfont\def\encodingdefault{LGR}}
\DeclareRobustCommand{\textgreek}[1]{\leavevmode{\greektext #1}}
\ProvideTextCommand{\~}{LGR}[1]{\char126#1}

\newcommand{\lyxmathsym}[1]{\ifmmode\begingroup\def\b@ld{bold}
  \text{\ifx\math@version\b@ld\bfseries\fi#1}\endgroup\else#1\fi}

\@ifundefined{textcolor}{}
{%
 \definecolor{BLACK}{gray}{0}
 \definecolor{WHITE}{gray}{1}
 \definecolor{RED}{rgb}{1,0,0}
 \definecolor{GREEN}{rgb}{0,1,0}
 \definecolor{BLUE}{rgb}{0,0,1}
 \definecolor{CYAN}{cmyk}{1,0,0,0}
 \definecolor{MAGENTA}{cmyk}{0,1,0,0}
 \definecolor{YELLOW}{cmyk}{0,0,1,0}
}

\makeatother

\graphicspath{{./images/}}

\begin{document}

\title{Recent Progress in Proximity Coupling of Magnetism to Topological
Insulators}
\author{Semonti Bhattacharyya}
\altaffiliation{Corresponding author: semonti.bhattacharyya@monash.edu and michael.fuhrer@monash.edu}
\affiliation{School of Physics and Astronomy, Monash University, Victoria 3800, Australia}
\affiliation{ARC Centre of Excellence in Future Low-Energy Electronics Technologies,
Monash University, Victoria 3800 Australia}

\author{Golrokh Akhgar}
\affiliation{School of Physics and Astronomy, Monash University, Victoria 3800, Australia}
\affiliation{ARC Centre of Excellence in Future Low-Energy Electronics Technologies,
Monash University, Victoria 3800 Australia}
\author{Matt Gebert}
\affiliation{School of Physics and Astronomy, Monash University, Victoria 3800, Australia}
\affiliation{ARC Centre of Excellence in Future Low-Energy Electronics Technologies,
Monash University, Victoria 3800 Australia}
\author{Julie Karel}
\affiliation{ARC Centre of Excellence in Future Low-Energy Electronics Technologies,
Monash University, Victoria 3800 Australia}
\affiliation{Department of Materials Science and Engineering, Monash University, Clayton, VIC 3800, Australia}
\author{Mark T Edmonds}
\affiliation{School of Physics and Astronomy, Monash University, Victoria 3800, Australia}
\affiliation{ARC Centre of Excellence in Future Low-Energy Electronics Technologies,
Monash University, Victoria 3800 Australia}
\author{Michael S Fuhrer}
\altaffiliation{Corresponding author: semonti.bhattacharyya@monash.edu and michael.fuhrer@monash.edu}
\affiliation{School of Physics and Astronomy, Monash University, Victoria 3800, Australia}
\affiliation{ARC Centre of Excellence in Future Low-Energy Electronics Technologies,
Monash University, Victoria 3800 Australia}

\begin{abstract}

Inducing long-range magnetic order in three-dimensional topological insulators can gap the Dirac-like metallic surface states, leading to exotic new phases such as the quantum anomalous Hall effect or the axion insulator state. These magnetic topological phases can host robust, dissipationless charge and spin currents or unique magnetoelectric behavior, which can be exploited in low-energy electronics and spintronics applications. Although several different strategies have been successfully implemented to realize these states, to date these phenomena have been confined to temperatures below a few Kelvin. In this review, we focus on one strategy, inducing magnetic order in topological insulators by proximity of magnetic materials, which has the capability for room temperature operation, unlocking the potential of magnetic topological phases for applications. We discuss the unique advantages of this strategy, the important physical mechanisms facilitating magnetic proximity effect, and the recent progress to achieve, understand, and harness proximity-coupled magnetic order in topological insulators. We also highlight some emerging new phenomena and applications enabled by proximity coupling of magnetism and topological materials, such as skyrmions and the topological Hall effect, and we conclude with an outlook on remaining challenges and opportunities in the field.

\end{abstract}

\maketitle


\section{Introduction}
\label{section: Introduction}

The quantum Hall effect (QHE) is an extraordinary materials phenomenon: a two-dimensional semiconductor in a high magnetic field can exhibit a  Hall conductivity quantized with part-per-billion precision in integer multiples of only fundamental constants $(e^{2}/h)$~\citep{klitzing1980new, von1986quantized, von2017quantum,  RN1225,thouless1982quantized}. The QHE state is accompanied by \emph{zero} longitudinal resistivity, and can carry current over macroscopic distances with no dissipation. The QHE state is now understood to be an example of a topological insulator~\citep{thouless1982quantized}, distinguished by the topology of its filled electron bands. The QHE state has even been realized at room temperature~\citep{novoselov2007room}, however requires impractically high magnetic fields. The question arises, can useful topological states be engineered in materials without high magnetic fields? 

Interest in this question has exploded since the discovery of time-reversal-symmetric topological insulators: the quantum spin Hall insulator (QSHI) in two dimensions (2D)~\citep{kane2005z,konig2007quantum} and the strong topological insulator in three dimensions (3D)~\citep{fu2007topological,hsieh2008topological,hsieh2009tunable,hsieh2009observation} to which here we will refer simply as ``topological insulator" (TI)~\citep{RN874, doi:10.1146/annurev-conmatphys-062910-140432}. The non-trivial topological band structures of these materials arise from an inversion of the normal order of bulk bands due to spin-orbit coupling, which gives rise to topologically protected boundary states. The one-dimensional (1D) edges of the QSHI and 2D surfaces of the TI are free from backscattering by non-magnetic disorder, cannot be localized, and generically possess linear (Dirac) dispersions. The characteristic energy scales for spin-orbit coupling can be large in heavy elements, resulting in inverted bandgaps much greater than room temperature in some cases~\citep{reis2017bismuthene, doi:10.1146/annurev-conmatphys-062910-140432, RN874}, bringing topological behaviour to room temperature without magnetic fields. 

However, the topological protection against backscattering in QSHIs and TIs is fragile, and can be destroyed by magnetic impurities, interaction with conduction electrons, or even inelastic scattering~\citep{liu2020helical}. The introduction of magnetic order to a topological material is a powerful way to realize new topological phases with more robust topological properties~\citep{haldane1988model}. In particular, establishing magnetic order in a QSHI or thin TI can produce a quantum anomalous Hall insulator (QAHI) with chiral 1D edges which are protected by the large energy scale of the inverted bandgap, and should be as robust to backscattering as the QHE~\citep{liu2008quantum,yu2010quantized}. This effect is known as quantum anomalous Hall effect (QAHE). The energy scale for magnetic order (the exchange energy) may also greatly exceed room temperature in common ferromagnets, hence robust topologically protected zero resistance materials at room temperature seem a realistic possibility. The QAHI is also a source of 100\% polarized spin currents, potentially controlled by electric fields and/or magnetization direction, with applications to spintronics~\citep{yasuda2017quantized, RN901, RN902, RN904}. Combining magnetic order and topology also allows entirely new phases to be created, such as the axion insulator, with a strong and quantized coupling between electric and magnetic polarizations, also an attractive spintronics building block~\citep{Mogieaao1669, RN685}. The QAHI and axion insulator states are briefly reviewed below in Section~\ref{subsection:Quantum anomalous Hall insulator (QAHI) state and axion insulator state}.

A number of strategies have been adopted to address this grand challenge of engineering robust topological and magnetic order in a materials system~\citep{RN910,RN905,RN902,RN904,RN878,RN907}. In Section~\ref{section:Pathways to achieve magnetic ordering in topological insulator (TI)} we briefly review the various pathways to achieve magnetic order in TIs. In the remainder of this review we focus on one strategy in particular: coupling of a TI to a magnetic material (MM) across an interface. This scheme has an important advantage: It enables separate optimization of the TI and MM properties, in particular the choice of materials with large inverted bandgap (TI) and high magnetic ordering temperature. However, this involves a significant challenge in that it relies on strong coupling of TI and MM by proximity at the interface. This review will discuss the various mechanisms for proximity-induced magnetism at the MM--TI interface, and review the recent progress in this field. We will highlight emergent phenomena made possible by MM--TI coupling, and prospects for applications. We will conclude with a perspective on remaining challenges and opportunities in this field. 
Whilst several other reviews~\citep{RN910,RN905,RN902,RN904,RN878,RN901,RN907} and a book chapter~\citep{anthony2018molecular} have been published already on the topic of magnetic TIs (MTI) in general, and also discuss proximity coupled heterostructures, most of these reviews miss the significant recent progress in the field, and do not provide a perspective focused on proximity-coupled magnetism. Here we address the rapidly growing body of literature on proximity coupled MM--TIs, and provide a timely and critical assessment of progress in the field as well as remaining challenges and opportunities. While we note the seminal contributions to the field, we largely focus on recent progress (post-2017). 

Section~\ref{section: mechanisms of proximity} will summarize the possible mechanism of proximity coupling in MM--TI interfaces, \textit{i.e.} exchange coupling, Magnetic extension, and surface-state assisted magnetism. Next, progress in MM--TI interfaces will be reviewed in section~\ref{section:Recent results in topological insulator (TI) - magnetic material (MM) heterostructures}, covering in order the coupling of TIs to ferromagnetic metals (FMMs) (section~\ref{section:Ferromagnetic metal}), ferromagnetic insulators (FMIs) (section~\ref{section:ferromagnetic insulators}), dilute ferromagnetic semiconductors (DFMSs) (section~\ref{subsection:Dilute ferromagnetic semiconductor}), van der Waals ferromagnets (vdWFMs) (section~\ref{subsection:van der Waals layered ferromagnetic materials}), and antiferromagnets (AFMs) (section~\ref{section:Anti-ferromagnetic proximity effect}). Then, emergent phenomena in MM--TI structures will be reviewed in section~\ref{subsection:Emergent Phenomena}, including QAHI (section~\ref{subsubsection:Quantum Anomalous Hall Insulators}) and axion insulators (section~\ref{subsubsection: Axion Insulators}), as well as novel interface phenomena associated with real-space topological defects (skyrmions, topological Hall effect; section~\ref{subsubsection: Skyrmions}), and the possibility of magnon-mediated superconductivity (section~\ref{subsubsection:Magnon-mediated superconductivity}). Finally some perspectives for the future will be presented in section~\ref{section:Conclusion and Outlook} on the remaining challenges in the field of proximity-induced magnetism in TIs, and the potential for applications in electronic and spintronic devices.

\subsection{Quantum anomalous Hall insulator (QAHI) state and axion insulator state}
\label{subsection:Quantum anomalous Hall insulator (QAHI) state and axion insulator state}
\begin{figure*}[t]
\centering
\selectlanguage{english}%
\includegraphics[width=\textwidth]{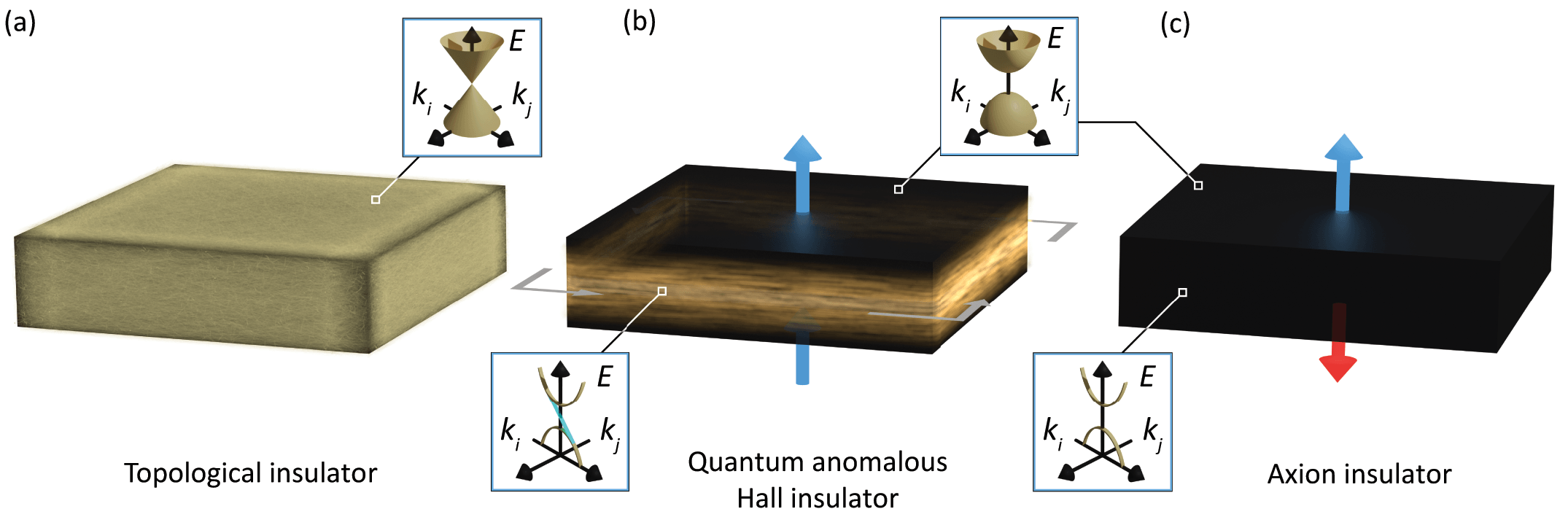}

\selectlanguage{british}%
\caption{\foreignlanguage{english}{Emergent phenomena in magnetized topological insulators (TIs). a) A 3D topological insulator (3D TI) has insulating bulk and 2D metallic surface states with gapless Dirac dispersion (inset). Magnetization of the 3D TI surfaces gaps the surface dispersion (inset), and leads to (b) quantum anomalous Hall insulator (QAHI) or (c) axion insulator states.  Blue and red arrows indicate the exchange field orientation; grey arrows indicate the chirality of the edge state. Additional insets show the dispersions of the chiral gapless (QAHI) and gapped (axion insulator) edges.
\label{fig:Emergent-phenomena}}}
\end{figure*}

Figure~\ref{fig:Emergent-phenomena} shows conceptual illustrations of a 3D TI (Figure~\ref{fig:Emergent-phenomena}a), a QAHI (Figure~\ref{fig:Emergent-phenomena}b) and an axion insulator (Figure~\ref{fig:Emergent-phenomena}c). 3D TIs have a unique electronic structure; the bulk of the material resembles a semiconductor where the conduction band and the valence band are separated by a band gap, while gapless spin-polarized surface states spanning the bulk band gap with a graphene-like linear dispersion relation appear as a result of the non-trivial topology of the bulk bands. The spin-momentum coupling of these surface states protects them from 180\degree~back-scattering and localization. The
topology of the band structure ensures the robustness
of the surface state unless some fundamental symmetry e.g.
time-reversal symmetry, or gauge symmetry is broken. If time-reversal symmetry is broken at the surface of a TI, QAHI (Figure~\ref{fig:Emergent-phenomena}b) and axion insulator (Figure~\ref{fig:Emergent-phenomena}c) states can emerge. When a surface of a 3D TI is magnetized in the out-of the plane direction, a magnetic exchange gap opens at the Dirac point of the surface state. Surprisingly, when the Fermi energy lies in the gap, the insulating surface is associated with a non-zero Hall conductance of quantized value $\pm\frac{e^2}{2h}$~\citep{yu2010quantized,chang2013experimental}. The sign of the Hall conductance is determined by the direction of the magnetization. This has a deep consequence in determining whether a TI with broken symmetry on both surfaces results in a QAHI or an axion insulator phase. Here, the side surfaces of the TI are not magnetized, and therefore have no magnetic energy gap. However, if the TI is very thin, the side surfaces may be gapped due to confinement by the gapped top and bottom surfaces.

As illustrated in Figure~\ref{fig:Emergent-phenomena}b), when both the surfaces have  magnetizations parallel to each other and perpendicular to the TI film, the half-integer quantum Hall conductances from each surface add, resulting in a total Hall conductance $\pm(\frac{e^2}{2h}+\frac{e^2}{2h}=\frac{e^2}{h})$~\citep{chen2010massive,chang2013experimental,mogi2015magnetic}. As long as the edges are gapped as discussed above, this phase will exhibit the integer QHE without a magnetic field, in other words, a quantized version of the anomalous Hall effect, the QAHE~\citep{haldane1988model}. Here, the chirality, hence the sign of the Hall conductance, is determined by the common magnetization direction. On the other hand, when the magnetization of the two surfaces are antiparallel, the half-integer quantum Hall conductance contributions from both channels cancel each other, resulting in $(\frac{e^2}{2h}-\frac{e^2}{2h}=0)$ Hall conductance (Figure~\ref{fig:Emergent-phenomena}c) as well as zero longitudinal conductance. However this phase differs from the usual insulator in that it retains an unusual quantized magnetoelectric coupling; this is known as the axion insulator phase~\citep{RN904,RN912,Mogieaao1669}. 

While both the QAHI and axion insulator phases require magnetization, both can be realized without external magnetic field. Thus the unique properties of these phases are fundamentally \emph{materials} properties, and can be exploited without resorting to high external magnetic fields. The QAHE features chiral edge states which can carry dissipation-free currents over macroscopic distances, which could be exploited for low-power interconnects. Furthermore the QAHI edges carry perfectly spin-polarized currents whose polarization direction depends on the magnetization, an attractive aspect for spintronics. Electrical control of the QAHE via electrostatic gate could lead to new types of low energy switches of charge and spin currents~\citep{RN688, Houeaaw1874}. Apart from this, the possibility of manipulation of chiral edge conduction at magnetic domain walls~\citep{yasuda2017quantized}, make them also a promising candidate for low-power spintronics and topological quantum computing. These aspects are explored further in Section~\ref{subsection:Future applications}.

As shown in Figure~\ref{fig:Emergent-phenomena}c), the merit of an axion insulator is that it suppresses all longitudinal electrical conductivity of the surfaces and edges, but retains the topological nature of the bulk TI. The absence of longitudinal conduction unveils the topological magnetoelectric effect (TME)~\citep{wilczek1987two,RN912,RN874,RN622,Mogieaao1669,RN685,qi2011topological}, an unusual coupling between electric \textbf{E} and magnetic \textbf{B} fields in TIs, which is otherwise shorted out by conducting surfaces which prevent electric polarization. The TME is a quantised version of magnetoelectric effect where external electric (\textbf{E}) and magnetic fields (\textbf{B}) are expected to cross induce magnetization (\textbf{M}) and electric polarization (\textbf{P}) respectively. This cross-induction can be expressed as

\begin{equation} \label{eu_eqn}
\mathbf{P}=\alpha/4{\pi}\mathbf{B}
\end{equation}

\begin{equation} \label{eu_eqn}
\mathbf{M}=-\alpha/4{\pi}\mathbf{E}
\end{equation}

where, $\alpha = \frac{1}{4\pi\varepsilon}\frac{e^2}{\hbar}$, the magnetoelectric susceptibility is the fine structure constant~\citep{qi2008topological,RN912,morimoto2015topological, RN874, RN622}. Hence, the TME can be used to measure $\alpha$ directly. More importantly, magnetoelectric coupling can provide a fundamental ingredient in spintronics, and may find other applications in energy transformation, signal generation and processing, information storage. These aspects are also explored further in Section~\ref{subsection:Future applications}. 
 
\subsection{Pathways to achieve magnetic ordering in topological insulators (TIs)}
\label{section:Pathways to achieve magnetic ordering in topological insulator (TI)}

To date, the QAHI and the axion insulator phases have been pursued in three types of magnetic topological insulator (MTI) systems: a) dilute magnetically doped TIs, b) intrinsic MTIs, c) magnetic material-topological insulator (MM--TI) heterostructures. In this section we will briefly describe the first two approaches before introducing the third strategy which is the focus of this review.

\subsubsection{Dilute magnetically doped topological insulators (TIs)}

\begin{figure*}[t]
\selectlanguage{english}%
\includegraphics[width=\textwidth]{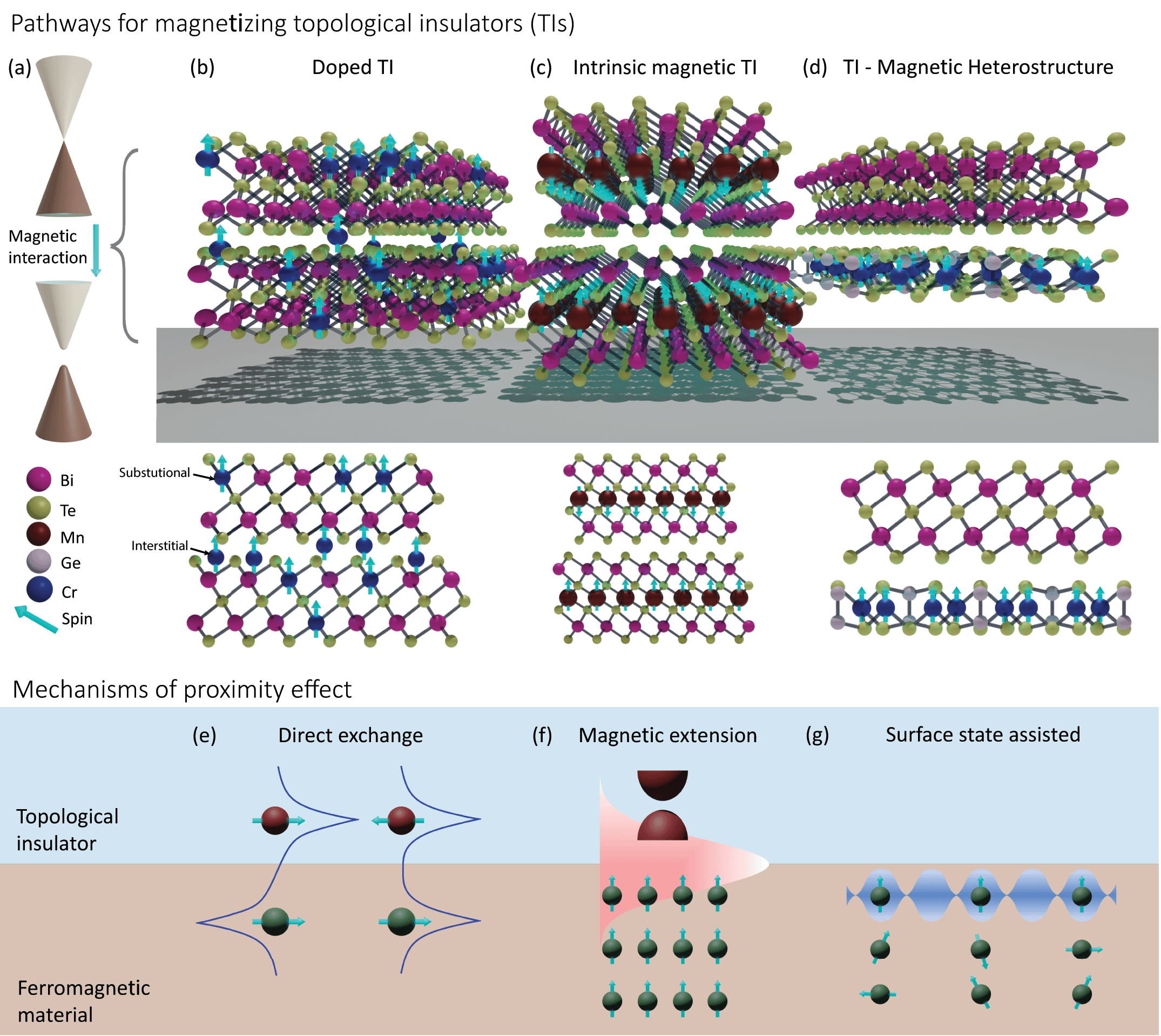}

\selectlanguage{british}%
\caption{Pathways to achieve magnetic ordering in topological insulators (TIs) and mechanisms of magnetic proximity effect. a) Magnetic interaction opens an exchange gap in the topological surface states (TSS) of a TI. Crystal structures of (b) a dilute magnetically doped TI (e.g. (Cr$_{x}$,Bi$_{1-x}$)${_2}$Te${_3}$), (c) an intrinsic magnetic TI (e.g. MnBi${_2}$Te${_4}$), and (d) a magnetic material (MM)--TI heterostructure e.g. Bi${_2}$Te${_3}$--Cr$_{2}$Ge$_{2}$Te$_{6}$. Microscopic mechanisms for magnetic proximity effect at a MM--TI interface: (e) direct exchange interaction,
(f) Magnetic extension, (g) TSS-mediated exchange coupling.}\label{fig:Pathways-to-achieve}
\end{figure*}

\selectlanguage{english}%
Figure~\ref{fig:Pathways-to-achieve}b) schematically illustrates the concept of dilute magnetically
doped TIs. In these materials, \emph{3d} transition metal ions
like Co, Cr, V, Fe, or Mn are embedded in subsitutional or
interstitial sites in the topological insulator crystals of 
$\mathrm{Bi}{}_{2}\mathrm{Se}{}_{3}$, $\mathrm{Bi}{}_{2}\mathrm{Te}{}_{3}$,
$\mathrm{(Bi,Sb)}{}_{2}\mathrm{Te}{}_{3}$ {etc}~\citep{yu2010quantized,RN878}.
The partially filled \textit{3d} electron shells of the transition
metal atoms host isolated magnetic moments. When such atoms
are introduced into the TI crystal, the local moments in these atoms can form long
range magnetic order through various possible mechanisms such as the
carrier-mediated Ruderman--Kittel--Kasuya--Yosida (RKKY) mechanism~\citep{Magnetic_impurities_on_the_surface_liu_2009magnetic,Ordering_of_magnetic_impurities_abanin2011ordering,Electrically_controllable_surface_magnetism_zhu2011electrically}, the local valence-electron-mediated Bloembergen--Rowland mechanism~\citep{Electrically_controllable_surface_magnetism_zhu2011electrically}
or the van Vleck mechanism~\citep{yu2010quantized,chang2013experimental,Thin_films_of_magnetically_doped_chang2013thin,Experimental_verification_of_the_Van_Vleck_li2015experimental}.
Inspired by the well-established field of dilute magnetic semiconductors,
experimentalists started exploring this method in Bi$_{2}$Se$_{3}$-family of 3D TIs in 2010~\citep{chen2010massive,Development_of_ferromagnetism_hor2010development}
 soon after their discovery in 2009~\citep{TI_theory_zhang2009topological,Bi2Se3_TI_observation_xia2009observation,Bi2Te3_observation_chen2009experimental} and the prediction of the QAHE in such systems~\citep{yu2010quantized}. This approach is the most widely explored technique to
synthesize magnetic topological insulators (MTIs) so far, and several review
articles have covered this topic in detail~\citep{RN902,RN903,RN878,RN901,RN904,RN905,RN907}.

The conditions generally considered necessary to achieve the QAHI phase  with dilute magnetic dopants Ti, V, Cr, or Mn are given below~\cite{yu2010quantized}: 

(i) The film should have a long-range ferromagnetic order with perpendicular magnetic anisotropy established through magnetic interactions in an insulating state. Yu \emph{et al.}~\cite{yu2010quantized} showed, in the case of the Bi$_{2}$Se$_{3}$-group of materials this is successfully achieved, because carrier-independent van Vleck susceptibility is significant due to the inverted bulk band structure of the TIs. Cr and Fe were found to be the most suitable dopants as these two dopants do not give rise to an impurity state in the bulk band gap.

(ii) The film should be thin enough to localise the surface and bulk transport channels, but thick enough to ensure the surface bandgap induced by hybridization of the two surfaces is less than the ferromagnetic exchange gap.

(iii) The chemical potential of the whole system should be fine-tuned to be within the surface band gap.

The first experimental demonstration of the QAHE was obtained in Cr-doped (Bi$_{0.1}$Sb$_{0.9}$)$_{1.8}$Te$_3$ samples~\citep{chang2013experimental}. Later, QAHE was also demonstrated in V-doped (Bi$_{1-y}$Sb$_{y}$)$_{2}$Te$_{3}$~\citep{V_doped_QAHE_chang2015high}. As the axion insulator requires a non-uniform magnetization of the crystal, it is not possible to obtain an axion insulator state in uniformly doped dilute MTIs~\cite{morimoto2015topological}. 

Despite the early success, dilute magnetic doping has
some inherent drawbacks. The dopant atoms occupy either substitutional
sites~\citep{Stability_electronic_and_magnetic_propertieszhang2013stability,Study_of_the_structural_electric_andwatson2013study,figueroa2014magnetic,liu2014local,sessi2016dual},
or interstitial sites~\citep{Study_of_the_structural_electric_andwatson2013study,liu2014local} (Figure~\ref{fig:Pathways-to-achieve}b),
leading to disorder and degradation of the quality of the crystal, which in turn
introduces defect states in the band gap of the TI~\citep{sessi2016dual,krieger2017spectroscopic,bhattacharyya2016resistance},
or causes unintentional high doping~\citep{xu2012hedgehog,tung2017tuning}. Sometimes, these samples are also prone to have inhomogeneity issues~\citep{lachman2015visualization,Lee1316,RN1101},
which result in microscopic phase separation, and prevent true long-range
ordering of the system. All these can 
result in compromised electrical transport performance, and confine the QAHE to low temperatures and low currents~\citep{gotz2018precision,pan2020probing}. To date, the highest temperature at which the QAHE is observed in uniformly-doped systems is 300~mK~\cite{doi:10.1002/adma.201703062}, despite the much higher ferromagnetic $T_{c}$ ($\sim$170~K)~\citep{zhou2005thin,V_doped_QAHE_chang2015high}, and ferromagnetic gap (few 100s of K)~\cite{Lee1316} (though modulation doping--a form of heterostructuring--has achieved QAHI at higher temperatures~\citep{Mogieaao1669} as discussed below). Because of all these reasons, new advances using this approach appear to have slowed.

\selectlanguage{british}%

\subsubsection{Intrinsic magnetic topological insulators (MTIs)}

\selectlanguage{english}%
As mentioned above, it appears increasingly more difficult to enhance the QAHE temperature in dilute magnetically doped TIs. This has caused researchers to seek other strategies. One such strategy is to search for intrinsic magnetic TI materials where magnetic elements are ordered within the crystal lattice~\citep{mong2010antiferromagnetic}. Instead of introducing magnetic impurities into TIs, this approach relies on synthesizing new stoichiometric
crystals which are intrinsically ferromagnetic and topological. 

In 2019, researchers theoretically predicted and experimentally confirmed that layered material MnBi$_2$Te$_4$ (first synthesized in 2013~\citep{lee2013crystal}) is a topologically non-trivial antiferromagnetic (AFM) TI in the bulk~\citep{otrokov2019prediction, li2019intrinsic,zhang2019topological,PhysRevX.9.041040,lee2019spin,vidal2019surface,zeugner2019chemical}. Figure~\ref{fig:Pathways-to-achieve}c) shows the atomic crystal structure of MnBi${_2}$Te${_4}$. MnBi${_2}$Te${_4}$ is a layered van der Waals (vdW) structure with atoms bonded covalently in  septuple layer (SL) units (Te-Bi-Te-Mn-Te-Bi-Te); it can be thought of conceptually as the insertion of a magnetic Mn-Te layer into the TI Te-Bi-Te-Bi-Te layer of Bi${_2}$Te${_3}$~\citep{li2019intrinsic,zhang2019topological}. ARPES measurements on MnBi${_2}$Te${_4}$ have shown heavy n-type doping with the fermi level lying in the bulk conduction band~\citep{otrokov2019prediction,PhysRevX.9.041040}. The magnetic moments of the Mn atoms in each SL of MnBi${_2}$Te${_4}$ point out of plane and are ferromagnetically coupled.  On other hand, the magnetic moments of two adjacent SLs tend to align antiparallel, forming A-type AFM states in the bulk of MnBi${_2}$Te${_4}$~\citep{otrokov2019prediction,li2019intrinsic,zhang2019topological}. As a consequence, MnBi${_2}$Te${_4}$ has gathered tremendous attention  since 2019 in the quest for QAHI due to its topological band structure and spontaneous magnetization without any external magnetic element doping~\citep{chowdhury2019prediction,otrokov2019unique,li2019intrinsic,zhang2019topological}.

Researchers have predicted that the single SL of MnBi${_2}$Te${_4}$ is a trivial 2D FMI and the topological and magnetic nature of this
material can be tuned by varying the thickness~\citep{li2019intrinsic}, with axion insulator occurring in even number of SLs and QAHI in odd number (greater than 1) of SLs
~\citep{zhang2019topological,li2019intrinsic,yu2019magnetic,otrokov2019unique}. The reason that the QAHE can survive in odd-numbered SLs of MnBi${_2}$Te${_4}$ is that the magnetic moments of all SLs but one compensate each other, yielding an uncompensated anti-ferromagnetic state~\citep{hao2019gapless}. This material has been successfully synthesized by both molecular beam epitaxy (MBE)~\citep{gong2019experimental} and single crystal growth~\citep{zhang2019experimental}. Both the QAHI and axion insulator state have been observed in this material below 2~K~\citep{deng2020quantum, liu2020robust,trang2020crossover}. (Another Chern insulator phase with zero longitudinal resistance and quantized Hall resistance $h/e^{2}$, has been reported in this material at high magnetic field and at temperatures up to 6.5~K~\citep{deng2020quantum, liu2020robust}; as this requires high magnetic field and formation of Landau levels, this is distinct from the zero-field QAHE.)

The realization of the QAHE and axion insulator phases in the intrinsic MTI MnBi${_2}$Te${_4}$ is an important breakthrough. However, the low N\'{e}el temperature ($\sim{25}$~K) is a barrier to non-cryogenic applications and indicates that other intrinsic MTI systems with higher ordering temperatures will need to be identified. This field was recently reviewed by Fei et al. in~\citep{RN910}, and is still rapidly evolving. As the main focus of the present review is proximity coupled magnetism, we will not discuss intrinsic MTIs further, except to comment below on the prospects for marrying proximity induced magnetization with intrinsic MTIs in heterostructures~\citep{otrokov2017highly}.

Intrinsic MTIs could also be interfaced with other materials, combining this approach with the proximity coupling approach (discussed below). Particularly, MTIs have recently been realized in superlattices based on Bi$_{2}$Te$_{3}$ and MnBi${_2}$Te${_4}$~\citep{aliev2019novel,RN649}, with topological properties confirmed to be non-trivial through theoretical calculations~\citep{hu2020van,vidal2019topological,sun2019rational}. Such materials may be thought of as either novel stoichiometric MTIs, or an infinite series of MM--TI heterojunctions. ARPES measurements have also been carried out by several groups~\citep{wu2019natural,hu2020van,vidal2019topological}. They have claimed that MnBi$_{4}$Te$_{7}$ is a new intrinsic MTI as it was predicted to realise the QAHE with much lower B field than MnBi${_2}$Te${_4}$~\citep{wu2019natural,hu2020van}. While this approach is interesting, and shares commonalities with the proximity coupling approach, we leave it outside the scope of this review.

\selectlanguage{british}%
\subsubsection{Magnetic material (MM)--Topological insulator (TI) heterostructures}

\selectlanguage{english}%
The remainder of this review focuses on proximity coupling of TIs and MMs at an interface. As illustrated in Figure~\ref{fig:Pathways-to-achieve}d), this particular methodology comprises creating a heterostructure of a TI and a MM, e.g. a ferromagnetic, ferrimagnetic, or antiferromagnetic
metal or insulator. The MM induces magnetic order in the TI
through proximity coupling. As a specific example, we have shown a heterostructure of the van der Waals ferromagnetic insulator (vdWFMI) Cr$_{2}$Ge$_{2}$Te$_{6}$ and the TI Bi$_{2}$Te$_{3}$ in Figure~\ref{fig:Pathways-to-achieve}d). This technique was inspired by the success
of proximity coupling of magnetism and superconductivity~\citep{sarma1963influence,de1966coupling,hao1991thin}. As the magnetic proximity effect has a relatively short length scale (few {\AA}), time-reversal symmetry is broken only at the interface of the TI and MM, and not in the bulk of the TI. Hence in order to achieve QAHE or axion insulator state, the TI must be sandwiched in between two layers of magnetic insulator (MI) with perpendicular magnetic anisotropy, \emph{i.e.} magnetization direction perpendicular to the TI film~\cite{Houeaaw1874}. 

This technique offers some unique advantages over the two methods
mentioned earlier. In \foreignlanguage{british}{dilute
magnetically doped TIs} or \foreignlanguage{british}{intrinsic
MTIs}, the TSS, and the magnetism, the two ingredients required for observing a QAHI or axion insulator, are properties of the same material, which makes it challenging to simultaneously
optimize both. On the other hand, in a MM--TI heterostructure system, TSS
and magnetism arise as properties of two different materials, allowing independent
control of both, e.g. the magnitude of the inverted bandgap in the TI, and the magnetic ordering temperature in the MM, parameters which may be prohibitively difficult to optimize in a single material. This technique stands out in its versatility because of the sheer
number and variety of MMs that can be potentially proximitized
to TIs, and the diversity of the fabrication techniques
that can be explored; the recent progress in this field (mainly, 2017-present) is summarized in \Cref{tab:Details-of-measurements}.

Proximity coupling of TI and MM has an additional advantage which is not immediately obvious: As discussed in more detail in section~\ref{subsubsection:Surface-state-assisted magnetism}, the presence of the strongly spin-orbit coupled TI surface may \textit{enhance} magnetic order in the MM. Indeed there is tantalizing experimental evidence of significantly elevated $T_{c}$ in both the interfacial magnetism in TI
and the MM layer, even surpassing room temperature~\citep{Tange1700307,RN598,RN1078,RN1109}. 

In the next section we describe the mechanisms responsible for the proximity effect.

\selectlanguage{british}%

\section{Mechanisms of proximity effect}
\label{section: mechanisms of proximity}

Figures~\ref{fig:Pathways-to-achieve}e-g) illustrate schematically the magnetic proximity effect, which can be regarded as a permeation of the magnetic
ground state into a neighboring material through the interaction
between magnetic moments in the MM and the electron
spins in the neighboring material. It is important to understand the various
microscopic mechanisms underpinning the magnetic proximity effects
in MM--TI heterostructures. Mainly three types
of interactions have been considered for these kinds of heterostructures: direct exchange coupling~(Figure~\ref{fig:Pathways-to-achieve}e), magnetic extension~(Figure~\ref{fig:Pathways-to-achieve}f), and \selectlanguage{english}%
surface-state assisted magnetism~(Figure~\ref{fig:Pathways-to-achieve}g). \selectlanguage{british}%
We briefly discuss each of these mechanisms in the following sections.

\subsubsection{Direct exchange coupling}
\label{subsubsection:Direct exchange coupling}

Direct exchange coupling arises from the electrostatic interaction
of neighbouring electrons with wave function overlap as a consequence
of the Pauli exclusion principle. Figure~\ref{fig:Pathways-to-achieve}e)
illustrates the mechanism: As electrons
are fermions, the two-electron wavefunction consisting of a spatial
component and a spin component must be antisymmetric as a whole. If the bonding
(antibonding) states have lower eigenenergy, the spatial part of
the two-electron wavefunction is symmetric (antisymmetric) and the
spins are antiparallel (parallel), leading to an antiferromagnetic
(ferromagnetic) exchange interaction~\citep{simon2013oxford}.
The nature of the interactions acting on the electrons
dictate whether the antisymmetric or the symmetric overlap of the
spatial part of the wavefunction has lower energy and thus determines
whether the ground state is ferromagnetic or antiferromagnetic.

\selectlanguage{english}%
When a clean interface is obtained between a TI
and a MM, the magnetic moments in the MM can align spins
in parallel or antiparallel manner in the TI through
direct exchange interaction~\foreignlanguage{british}~{~\citep{RN693,RN588,RN592,RN647,RN664,RN973,RN589,RN1004}}.
This interaction requires direct wavefunction overlap and hence is inherently short ranged; thus for the TI surface state which decays exponentially into the TI bulk, the magnetization decays rapidly as well~\citep{RN598}. Similarly, only the magnetic moments of the layer
next to the MM have any relevant impact~\citep{RN978,RN611}.
Hence, even layered AFMs, where the intralayer exchange coupling is ferromagnetic and the interlayer exchange coupling is antiferromagnetic (A-type AFM)~\citep{mak2019probing}, can have the same proximity effect as ferromagnets~\citep{RN917,RN694,RN916}, as the TI will ``see" only the uniform ferromagnetically aligned spins of the interface layer. 

In addition to direct exchange, indirect exchange between the TSS and magnetic moments is also possible, mediated for example through
ordinary interface states created due to the sharpness of
the boundary~\citep{RN917}.
\selectlanguage{british}%

\subsubsection{Magnetic extension}
\label{subsubsection:Magnetic extension}

In the case of certain van der Waals ferromagnetic insulators (FMIs) like Cr$_{2}$Ge$_{2}$Te${_6}$ the magnetization induced in the electronic states of a TI in proximity by direct exchange is negligible and can not fully explain the observed exchange gap opened in the TI surface states~\citep{RN637}. Instead, a mechanism named ``magnetic extension" is used to explain the magnetization of the TI surface state. Figure~\ref{fig:Pathways-to-achieve}f) schematically explains this mechanism. The surface state of the TI decays exponentially into the bulk of the TI, but also decays into the FMI, where it interacts with magnetic moments via strong exchange interaction to produce a larger exchange gap at the Dirac point, compared to the case of ordinary proximity effect. It was proposed that this type of proximity effect would be more efficient in heterostructures
where the MM and TI have (i) exactly
the same or similar crystal structure, and (ii) similar atomic
composition~\citep{otrokov2017highly,RN686,RN699}. However, thus far, a limited range of structures have been explored theoretically, and ab initio calculations are difficult for non-commensurate structures, so it remains an open question how to optimize the heterostructures for magnetic extension. We will discuss this mechanism in detail in section~\ref{subsection:van der Waals layered ferromagnetic materials}, in the context of layered FMI--TI heterostructures.

\selectlanguage{english}%

\subsubsection{Surface-state-assisted magnetism}
\label{subsubsection:Surface-state-assisted magnetism}

Mounting experimental evidence suggests the presence of unique interactions
at the interface of TIs and magnetic insulators (MIs)~\citep{RN598,RN664,RN592} potentially without a counterpart in magnetic heterostructures involving topologically trivial materials. These include unusual
enhancement of $T_{c}$, canting of magnetic moments in a MI with natural in-plane magnetic
anisotropy~\citep{wei2013exchange,RN598}, and unexpected enhancement in the
magnitude of anomalous Hall effect~\citep{RN598,wei2013exchange}.
The origin of these effects are highly debated~\citep{RN693}, and
still being explored. However, several studies agree that this cohort of unique phenomena is associated with the non-trivial topology of the TI bands or
the spin-orbit coupling of the surface states, which too is a manifestation of
this non-trivial topology. One such proposed mechanism is surface-state-assisted enhancement of robustness of magnetic order at the interface, illustrated schematically in Figure~\ref{fig:Pathways-to-achieve}g). In this case, the highly-coherent TSS mediates an RKKY interaction between the interface magnetic moments of MI, enhancing its magnetic order and $T_{c}$~\cite{RN636,RN1125}. According to another proposed mechanism, perpendicular magnetic anisotropy is induced in FMI heterostructures due to strain caused by lattice mismatch and enhanced by the spin-orbit coupling (SOC) of the TSS~\cite{RN636}.

 Researchers have already started exploiting this strategy to engineer higher $T_{c}$ in magnetic materials~\citep{RN1078,RN1109}. A complete understanding of these type of mechanisms is necessary for successful manipulation of magnetic proximity effect in TI.

\section{Recent results in topological insulator (TI) - magnetic material (MM) heterostructures}
\label{section:Recent results in topological insulator (TI) - magnetic material (MM) heterostructures}

\subsection{Ferromagnetic proximity effect}
\label{section:Ferromagnetic proximity effect}

In the following section we will discuss (a) the recent advancements
in heterostructures of topological insulators (TIs) and ferromagnetic metals (FMMs; Section~\ref{section:Ferromagnetic metal}), (b) ferromagnetic insulators (FMIs; Section~\ref{section:ferromagnetic insulators}), (c) dilute ferromagnetic semiconductors (DFMSs; Section~\ref{subsection:Dilute ferromagnetic semiconductor}),
and (d) van der Waals ferromagnets (vdWFMs; Section~\ref{subsection:van der Waals layered ferromagnetic materials}). We will also include the proximity effect from ferrimagnetic material in this section, as although ferrimagnets and ferromagnets are different in microscopic details, below $T_{c}$ both possess net magnetization and can induce magnetic proximity effect through similar mechanisms.
\selectlanguage{british}%

\subsubsection{Ferromagnetic metal (FMM)}
\label{section:Ferromagnetic metal}
\selectlanguage{british}%
\begin{figure*}[t]
\selectlanguage{english}%
\includegraphics[width=\textwidth]{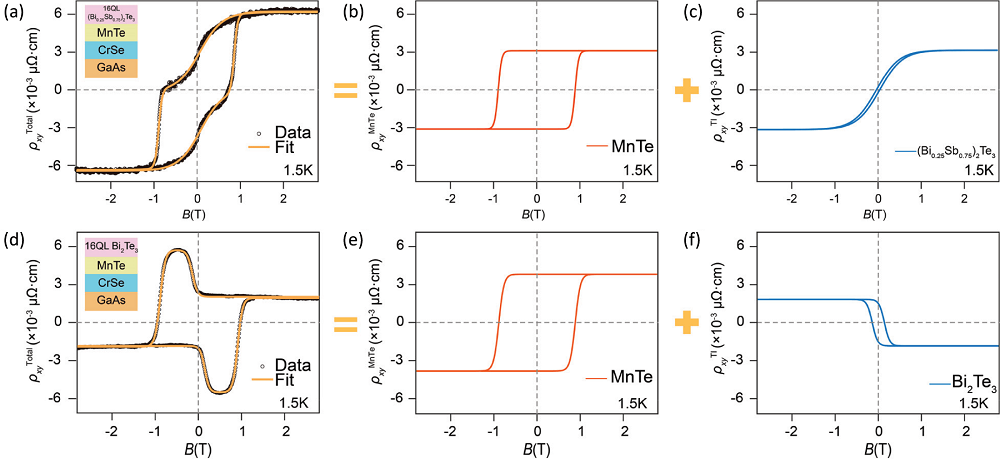}

\selectlanguage{british}%
\caption{\foreignlanguage{english}{Anomalous Hall effect (AHE) in ferromagnetic metal (FMM)--topological insulator (TI) heterostructures. Total Hall resistivity as a function of magnetic field $\mathrm{\rho}_{xy}$($B$) for (a) p-type 16 quintuple layer (QL) (Bi$_{0.25}$Sb$_{0.75}$)$_{2}$Te$_{3}$--MnTe heterostructure
and (d) n-type 16 QL Bi$_{2}$Te$_{3}$--MnTe heterostructure. The experimental data is represented by the black open dots, while the fitting results are indicated
by the khaki solid lines. The AHE data in (a) and (d) can
be decomposed into the contributions from MnTe (orange lines, (b) and (e) and TI (Bi$_{0.25}$Sb$_{0.75}$)$_{2}$Te$_{3}$
or Bi$_{2}$Te$_{3}$; blue lines in (c) and (f) respectively). Reprinted with permission from Nano Lett. 2020, 20, 3, 1731-1737~\citep{RN816}. Copyright (2020) American Chemical Society.\label{fig:TI-FM-metal-heterostructures.}}}
\end{figure*}

The study of FMM--TI heterostructures has
mainly been motivated by the prospect of efficient spintronic
devices, as TIs are known to generate high spin-orbit torque~\citep{fan2014magnetization,mellnik2014spin}. Various growth methods have been explored to synthesize such heterostructures, including laser molecular beam epitaxy (LMBE)~\citep{RN590,RN683},
molecular beam epitaxy (MBE)~\citep{RN593}, laser ablation deposition~\citep{RN627},
hot-wall epitaxy~\citep{RN591}, metal organic chemical vapour deposition
(MOCVD)~\citep{RN599}, radio-frequency (RF) magnetron sputtering~\citep{RN642}
etc. Even though proximity coupling has been successfully achieved in FMM--TI heterostructures~\citep{vobornik2011magnetic,RN1090},
the FMM electrically shorts the TI channel, making it difficult to
detect any signature of magnetization in the TI in transport experiments~\citep{RN627}. In this section, we discuss electrical transport measurements in such heterostructures, with an emphasis on the techniques which may differentiate between the magnetizations of the TI, and the FMMs.

The anomalous Hall effect (AHE) provides an unambiguous signature of magnetization in electronic transport~\citep{nagaosa2010anomalous}, and is therefore one of the most widely used techniques to investigate MM--TI heterostructures. The dependence of the transverse resistivity $\rho_{xy}$ on magnetic field $H$ 
in ferromagnetic materials can be expressed as

\begin{equation}
\rho_{xy}(H)=\rho_{xy}^{O}(H)+\rho_{xy}^{A}(H)
\end{equation}

Here, the component {$\rho_{xy}^{O}(H)$ originates from the ordinary
Hall effect, which is a manifestation of the Lorentz force acting on the charge carriers. This component is also present in non-MMs, and is linearly
dependent on $H$ ($\rho_{xy}^{O}(H)=R_{0}H),\textrm{with }R_{0}$ being known as the ordinary Hall coefficient), and therefore $\rho_{xy}^{O}(H=0) = 0$.
$\rho_{xy}^{A}(H)$ is the anomalous component of $\rho_{xy}$, and
originates from broken time-reversal symmetry as a consequence of spin-orbit coupling.  The mechanisms that give rise to this effect can be classified into
two categories: (i) extrinsic mechanisms dependent on impurity scattering,
such as, skew scattering and side-jump; (ii) an intrinsic mechanism that
is dependent on the Berry curvature of electronic bands. Both of these
mechanisms exist only in the presence of exchange field and spin-orbit interaction
in the material. However, microscopic details of the materials dictate which mechanism is dominant. Note that the QAHE is a quantized version of the intrinsic AHE, due to the quantized integral of the Berry curvature over the full band of the QAHI.

In particular, a non-zero $\rho_{xy}^{A}(H = 0)$ is indicative of a magnetization. For many materials, it is observed that $\rho_{xy}^{A}$(H) is proportional to the magnetization $M$(H)}

\selectlanguage{english}%
\begin{equation}
\rho_{xy}^{A}(H)=R_{A}M(H)
\end{equation}

where $R_{A}$ is known as the anomalous Hall coefficient and the resulting non-linear and hysteretic
behavior of  $\rho_{xy}^{A}(H)$ is known as the anomalous
Hall effect. As an experimental tool to detect the presence of a magnetization, it is generally assumed that a non-linear or hysteretic magnetic-field dependence of $\rho_{xy}^{A}(H)$ reflects the magnetization
of the electron transporting material.

Figure~\ref{fig:TI-FM-metal-heterostructures.} illustrates the use of the AHE to investigate FMM--TI heterostructures. Chen \emph{et.al.}~\citep{RN816} used the AHE to investigate TIs grown on single-crystalline FMM films of MnTe and provided a way to disentangle the magnetization and anomalous Hall signatures of the two components of such a heterostructure. Figure~\ref{fig:TI-FM-metal-heterostructures.}a) and d) show the anomalous Hall effect measured in MnTe--(Bi$_{0.25}$Sb$_{0.75}$)$_{2}$Te$_{3}$ and MnTe--Bi$_{2}$Te$_{3}$ heterostructures respectively. The device
schematics are shown in the inset. Whereas the AHE measurements $\rho_{xy}$($H$) on bare MnTe samples showed a square hysteresis loop of a typical FMM, additional steps or humps appear in $\rho_{xy}$($H$) for the heterostructure films. Similarly, the magnetic field-dependent magnetizations of these samples also showed a two-step transition feature with characteristic transition fields matching with the AHE transition fields (not shown).

The authors interpreted the AHE and the magnetization signatures as indicators of two coexisting magnetic phases (FMM and magnetized TI) with different magnetization reversal fields. The AHE was separated into signals from two independent magnetic phases applying the Weiss molecular field model simulations in Figure~\ref{fig:TI-FM-metal-heterostructures.}a-f)~\citep{mohn2003weiss}.
One of the components shown in Figure~\ref{fig:TI-FM-metal-heterostructures.}b) and e) is identified with MnTe as it is similar to the AHE for bare MnTe, while the other component, shown in  Figure~\ref{fig:TI-FM-metal-heterostructures.}c)
and f) is attributed to the magnetized TI. Remarkably, this AHE component has opposite
polarity for electron-doped Bi$_{2}$Te$_{3}$ (Figure~\ref{fig:TI-FM-metal-heterostructures.}f)
and hole-doped (Bi$_{0.25}$Sb$_{0.75}$)$_{2}$Te$_{3}$ (Figure~\ref{fig:TI-FM-metal-heterostructures.}c). The authors interpreted
this peculiar phenomena as the signature of the opposite
signs of the Berry curvature for different Fermi-level positions.  However, the Berry phase is neither expected to have opposite polarity for n- and p-type doping, nor has the polarity change been observed in other magnetized topological insulator systems such as, magnetically
doped TIs~\citep{kou2013interplay,RN907} or FMI--intrinsic MTI heterostructures~\citep{RN1005}. Hence, while the AHE and magnetization clearly probe two distinct magnetic phases, the origin of polarity reversal in the MnTe--TI system is not well understood, and may be an artifact arising from either the metallic nature of the MnTe substrate, or a disordered interface. 


\begin{figure}[!th]
\includegraphics[width=\columnwidth]{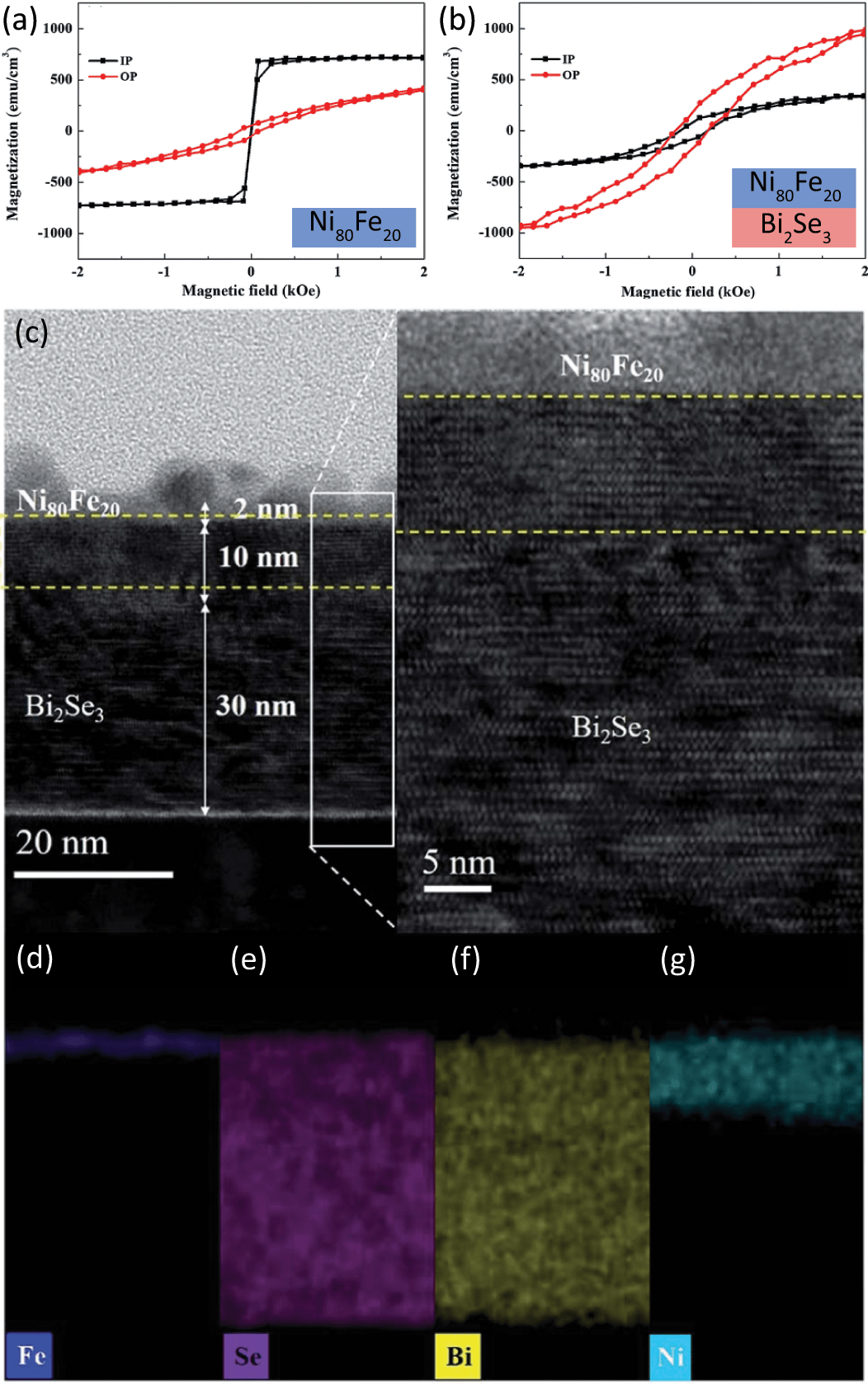}

\caption{\foreignlanguage{english}{Formation of intermediate phases in a ferromagnetic metal (FMM)--topological insulator (TI) heterostructure. a-b) Magnetization vs. magnetic field measured at 300 K, along the in-plane (IP) and out-of-plane (OP) directions for (a) pure FMM Ni$_{80}$Fe$_{20}$ permalloy (Py) and (b) heterostructure of Py and 20~nm Bi$_{2}$Se$_{3}$. All the magnetization curves were normalized to the sample area to ensure
a reasonable quantitative comparison. Insets of (a) and (b) show the
device schematic of pure Py and Py--Bi$_{2}$Se$_{3}$ respectively.
c) Cross-sectional transmission electtron microscopy (TEM) image of the Py--Bi$_{2}$Se$_{3}$ heterostructure. The inset in (c)
is an HRTEM image of the selected region of the heterostructure, in
which the dashed lines identify the intermediate phases with crystallinity
and lattice spacing different from those of amorphous Py (top) and
epitaxial Bi$_{2}$Se$_{3}$ (bottom).
d-g) Energy dispersive X-ray maps of (d) Fe, (e) Se, (f) Bi and (g) Ni taken in the interface region. Adapted from Ref.~\citep{RN593} - published by The Royal Society of Chemistry under CC BY 3.0 license (https://creativecommons.org/licenses/by/3.0/).}}
\label{fig:TI-FM-metal-heterostructures.-1}
\end{figure}


While the AHE can provide a powerful probe of magnetization of an electronic material, care must be taken in interpreting that magnetization as due to proximity effect. A material-dependent extrinsic origin of AHE, caused by
local doping or diffusion of ferromagnetic atoms can be a confounding issue. 
Indeed, diffusion of magnetic ions into TIs and formation of intermediate
phases at the interface has been one of the issues plaguing this field~\citep{RN591,RN599,RN642}.
For example, Chang \emph{et al.}~\citep{RN593} studied Permalloy (Py; Ni$_{80}$Fe$_{20}$) and Py-Bi$_{2}$Se$_{3}$ heterostructures. Figure~\ref{fig:TI-FM-metal-heterostructures.-1}a) and b) show the
magnetization vs. magnetic field data obtained for the two structures. Comparison of in-plane and out-of plane magnetization measurements in bare Py films reveals
a larger value of in-plane saturation magnetization, and smaller value of out-of-plane saturation magnetization, indicating strong in-plane
anisotropy and weak out-of-plane anisotropy. In contrast, the heterostructures
display weaker in-plane anisotropy. The out-of-plane anisotropy in heterostructure
was found to strengthen with thickness of the Ni$_{80}$Fe$_{20}$
layer. It seems attractive to attribute this modification of the magnetization behavior in the heterostructure
to proximity coupling of TI and FMM. However,
transmission electron microscopy and energy dispersive X-ray maps clearly shows the appearance of FeSe and Ni:Bi$_{2}$Se$_{3}$
phases at the interface (Figure~\ref{fig:TI-FM-metal-heterostructures.-1}c-g).
Synchrotron radiation photoelectron emission spectroscopy (SR-PES)
and X-ray absorption near edge structure (XANES) measurements corroborate
these findings as well. Further analysis of X-ray magnetic circular
dichroism spectra (XMCD) signals revealed that the total magnetization results from a complex interplay of weak perpendicular magnetic anisotropy
(PMA) of metallic FeSe, in-plane magnetic anisotropy (IMA) of insulating
Ni:Bi$_{2}$Se$_{3}$ and magnetic proximity on Bi$_{2}$Se$_{3}$.
This demonstrates that care must be taken in attributing magnetic signals to
proximity effect in such heterostructures~\citep{RN567,RN659}, and every new type of heteostructures prepared by a new technique
requires direct chemical scrutiny to eliminate the possibility of these type of artifacts.
\selectlanguage{british}%

\subsubsection{Ferromagnetic insulators (FMIs)}
\label{section:ferromagnetic insulators}

The use of FMIs in magnetic material--topological insulator (MM--TI) heterostructures rather than ferromagnetic metals (FMMs) can solve several problems. FMIs
do not short out the electrical transport signatures of TIs. Moreover, in FMI--TI heterostructures the chemical potential
of the TI can be tuned through the exchange gap using a FET geometry~\citep{Electrically_controllable_surface_magnetism_zhu2011electrically,RN688,RN647}, while the large density of states of a metal generally precludes tuning the chemical potential in a FMM--TI structure.
In addition, owing to the long-term research on FMIs in magnetic tunnel
barrier devices, protocols for growing these materials down to nm
thickness are well-established, enabling defect-free
smooth interfaces~\citep{moodera2007phenomena}. 

The FMIs utilized in these heterostructures generally need to fulfill several conditions
in order to make the induced exchange gap in TI observable to experiments: (i) small lattice parameter mismatch with the TI crystal is desirable to achieve optimal MBE-growth of the heterostructure, although films have been grown via van der Waals epitaxy on lattice mismatched substrates; (ii) chemically inert surfaces to avoid trivial Shockley
or Tamm interface states; (iii) proper band alignment of TI and FMI
which ensures that the charge neutrality point of TI appears within the FMI
band gap. A few examples of appropriate FMIs are lanthanide chalcogenides
(EuS)~\citep{wei2013exchange,RN598,RN636,RN688,RN630,RN597,RN692,RN693}, rare-earth garnets (e.g. yttrium iron garnet, YIG; thulium iron garnet, TIG)~\citep{Tange1700307,RN645,RN939,RN664}, and other ferromagnetic oxides~\citep{RN592,RN592}. Most works have studied FMIs with perpendicular anisotropy, however several heterostructures have been investigated, where the  FMIs originally has in-plane anisotropy~\cite{RN688,wei2013exchange,yang2013emerging}, as even in-plane magnetization can induce an exchange gap in the TI surface, provided reflection symmetry is broken~\cite{liu2013plane}. In the following subsections, we discuss several of these FMI--TI heterostructures, which are recently studied.

\paragraph{\textbf{EuS:}} EuS is a ferromagnetic semiconductor with a rocksalt-type
crystal structure. The half-filled $4f$-shell of Eu$^{2+}$ ($4f^{7}6s^{0}$)
ions is responsible for magnetism in these materials. These orbitals,
localized at Eu$^{2+}$ ion-sites, appear as degenerate levels within the band gap, and form
a Heisenberg magnet through exchange interaction~\citep{nolting1982modified}.
Bulk EuS has a Curie temperature $T_{c}=16.9$~K~\citep{RN986}. As EuS is a well-characterized FMI, and its growth protocols on various substrates are well-established~\citep{hao1991thin,moodera2007phenomena}, this is the first FMI
for which FMI--TI heterostructure was ever attempted~\citep{wei2013exchange,yang2013emerging}, demonstrating a clear signature of magnetic proximity effect and a large exchange gap of the TI surface states.

\begin{figure*}[t]
\selectlanguage{english}%
\includegraphics[width=\textwidth]{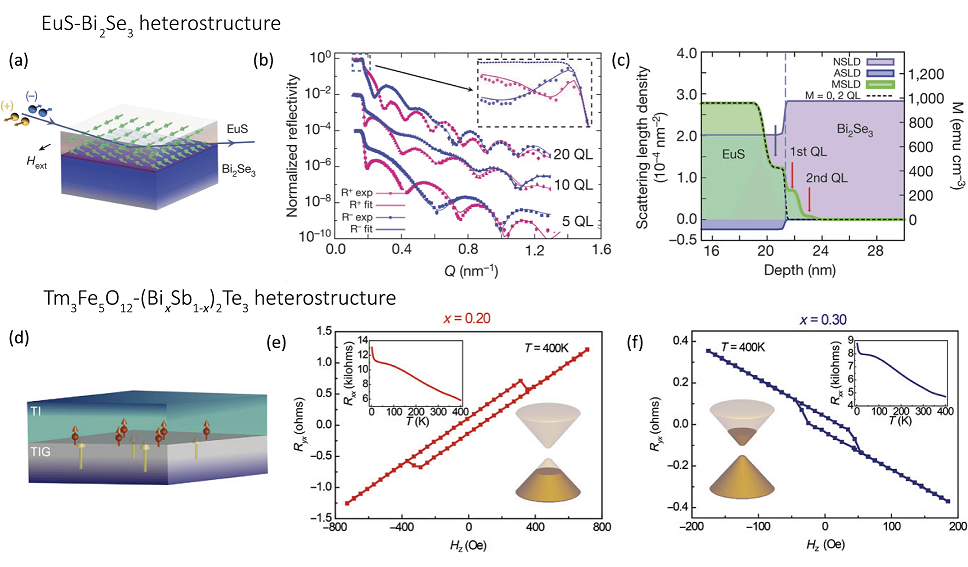}

\selectlanguage{british}%
\caption{\foreignlanguage{english}{Characterization of ferromagnetic insulator (FMI)--topological insulator (TI) heterostructures. a-c) Polarized neutron reflectivity (PNR) characterization of EuS--Bi$_{2}$Se$_{3}$ heterostructure. 
a) Schematic of the PNR experiment for a EuS--Bi$_{2}$Se$_{3}$ FMI--TI heterostructure.
b) Measured (symbols) and fitted (solid lines) reflectivity curves
for spin-up (R+) and spin-down (R\textminus ) neutron spin-states
shown on a logarithmic--linear scale as a function of momentum transfer
$\mathrm{Q}=4\pi\sin\left(\theta\right)/\lambda$, where $\theta$
is the incident angle and \textgreek{l} is the neutron wavelength.
The inset is an expanded view of the reflectivity below its critical
edge, where the reflectivity is sensitive to the distribution of the
Eu atoms owing to their absorption cross-section and their magnetic
moment. The error bars represent one standard deviation. c) PNR nuclear
(NSLD, in pink), magnetic (MSLD, in green) and absorption (ASLD, in
blue) scattering length density profiles, measured for EuS--Bi$_{2}$Se$_{3}$ (20 quintuple layer (QL)) at $5$~K with an external in-plane magnetic field of 1~T
and presented as a function of the distance from the sample surface.
The magnetization measured inside the Bi$_{2}$Se$_{3}$ layer is
marked with red arrows, and the reduction of the in-plane component
of EuS at the interface caused by a canting of the Eu magnetization
vector towards the out-of-the plane direction is marked with a blue arrow. The scale
on the right-hand-side shows the magnetization M inferred from the MSLD. The black dashed line is the fitted MSLD profile obtained assuming zero magnetization
(M=0 in 2 QL) in the Bi$_{2}$Se$_{3}$ layer. d-f) Anomalous Hall effect in FMI--TI heterostructure Tm$_{3}$Fe$_{5}$O$_{12}$ (TIG)--(Bi$_{x}$Sb$_{1-x}$)$_{2}$Te$_{3}$. d) Schematic of the FMI--TI heterostructure. e,f)
Hall resistance as a function of magnetic field $R_{yx}(H_{z})$ for TIG--(Bi$_{x}$Sb$_{1-x}$)$_{2}$Te$_{3}$ ($5$~QL)
for $x=0.20$ and $0.30$, respectively. The upper insets show the
corresponding temperature ($T$) dependence of the longitudinal resistivity $R_{xx}$. The lower
insets show schematic drawings of the corresponding chemical potential
positions within the TSS dispersions. \label{fig:Polarized-neutron-reflectivity} a-c) Reprinted by permission from RightsLink Printable License: Springer Nature Nature, A high-temperature ferromagnetic topological insulating phase by proximity coupling, Ferhat Katmis, Valeria Lauter, Flavio S. Nogueira, Badih A. Assaf, Michelle E. Jamer, Peng Wei, Biswarup Satpati, John W. Freeland, Ilya Eremin, Don Heiman, Pablo Jarillo-Herrero and Jagadeesh S. Moodera~\citep{RN598}, \copyright~Macmillan Publishers Limited.(2016) and d-f) adapted from Tang \emph{et al.}, Sci. Adv. 2017;3: e1700307~\citep{Tange1700307} under a Creative Commons Attribution NonCommercial License 4.0 (CC BY-NC) \url{http://creativecommons.org/licenses/by-nc/4.0/}}}
\end{figure*}

One of the important spectroscopic techniques utilized to study FMI--TI heterostructures is polarized neutron reflectometry (PNR). PNR provides a unique insight into
the detailed spatial dependence of vector magnetization inside the
bulk of the sample~\citep{majkrzak1991polarized}, as neutrons possess an intrinsic
magnetic moment, and neutrons with de Broglie wavelength comparable to the interatomic distances can be generated~\citep{majkrzak1991polarized}. In 2016, Katmis \emph{et al}.~\citep{RN598} measured depth
profile of magnetization in such structures ($20$~quintuple layer (QL) Bi$_{2}$Se$_{3}$ with a EuS capping layer possessing in-plane anisotropy) using PNR. Figure~\ref{fig:Polarized-neutron-reflectivity}a)
schematically shows PNR measurements in EuS--Bi$\mathrm{_2}$Se$\mathrm{_3}$ heterostructure.
In this technique, specular reflectivity is measured for neutrons
with magnetic moments parallel ($\textrm{R}+$) and antiparallel ($\textrm{R}-$) to the external magnetic
field as a function of momentum transfer $\textrm{Q}$ as shown in Figure~\ref{fig:Polarized-neutron-reflectivity}b).
For a magnetic layer with a non-magnetic substrate, the dependence of
$\textrm{R}\pm$ on $\textrm{Q}$ will be analogous to an optical Fresnel
diffraction pattern, where periodicity is determined by the neutron refractive
index of different layers and their respective thickness.
As the neutron refractive index depends on the relative orientation of the incoming neutron moments and the atomic moments in the sample, the dependence of \foreignlanguage{english}{$\textrm{R}+$
and $\textrm{R}-$ on $\textrm{Q}$ are different. By simultaneously measuring $\textrm{Q}$-dependence of both {$\textrm{R}+$} and $\textrm{R}-$, it is possible to determine thickness of thin films and the depth profile of their magnetic moments.} \selectlanguage{english}%

Figure~\ref{fig:Polarized-neutron-reflectivity}c) shows the resulting depth dependence
of nuclear scattering length density (NSLD) and magnetic scattering
length density (MSLD) obtained from a simultaneous fit to the reflectometry
data (Figure~\ref{fig:Polarized-neutron-reflectivity}b).
The NSLD profile reveals a sharp interface between the EuS and Bi$_{2}$Se$_{3}$
with average roughness $\sim0.2$~nm. On the contrary, MSLD shows
a more gradual dependence on depth, starting at $1.5$~nm inside EuS (blue
arrow in Figure~\ref{fig:Polarized-neutron-reflectivity}c), through
to first and second QL of Bi$_{2}$Se$_{3}$. The reduction of MSLD
in EuS is attributed to a canting of the Eu magnetization vector towards
the out-of-plane direction. However, the
magnetization of the first two QLs inside Bi$_{2}$Se$_{3}$ has
to be attributed to magnetic proximity effect, as NSLD measurements
and absorption scattering length density (ASLD) show that Eu atoms
are absent in Bi$_{2}$Se$_{3}$. Surprisingly, the magnetization
of Bi$_{2}$Se$_{3}$, albeit reducing with increasing temperature,
remains substantial at room temperature (not shown).
This extraordinary enhancement of $T_{c}$ can be considered as the
first step towards achieving practical applications of MTI.
As the $T_{c}$ ($16.9$~K) of EuS is far below room temperature, this remarkable
robustness of magnetization of Bi$_{2}$Se$_{3}$ perhaps indicates
a special role of topology in realization of this magnetic state.
\selectlanguage{british}%

The extraordinary behavior of the EuS--TI interface has remained
an active and highly debated topic of research~\citep{RN636,EREMEEV201530}.
Kim \emph{et al.}~\citep{RN636} studied these heterostructures
through systematic first principles calculations and model analysis. Contrary to the experiments which showed efficient proximity-driven magnetic induction at the interface layers of TI~\citep{wei2013exchange,RN1004},
the calculations found no induced magnetic moment. However, both perpendicular
magnetic anisotropy at EuS interface states and the enhancement of
$T_{c}$ were attributed to (i) high spin orbit coupling of TI, and
(ii) emergent interface Dirac states caused by charge redistribution
at the interface. Kim \emph{et al.}~\citep{RN636}
found that these Dirac states can extend into EuS and contribute to
additional exchange coupling of Eu magnetic moments through indirect
exchange interaction. On the contrary, Osterhoudt \emph{et al.}~\cite{RN630} found the usual EuS Raman mode to be absent in the heterostructure,
and attributed this absence to a transfer of charge between the EuS and the
Bi$_{2}$Se$_{3}$. Low energy muon spin rotation (LE-\textgreek{m}SR)
detected the presence of local magnetic fields extending from EuS into
the adjacent nonmagnetic layer for both EuS--Bi$_{2}$Se$_{3}$ and
EuS--titanium heterostructure, revealing a non-topological origin of the proximity-coupled magnetization~\citep{RN693}. It is important to emphasize that, in all the experimental studies mentioned above, Bi$_{2}$Se$_{3}$ is very highly doped ($10^{13}$-$10^{14}$~cm$^{-2}$), placing the chemical potential within or near bulk conduction band~\citep{analytis2010bulk,zhang2013weak}. Hence, the TSS may not have any involvement in the proximity effect. 

To address the issue of high doping, later experiments have concentrated on EuS - (Bi$_{x}$Sb$_{1-x}$)$_{2}$Te$_{3}$ heterostructure, as (Bi$_{x}$Sb$_{1-x}$)$_{2}$Te$_{3}$, is known
to have lower doping compared to Bi$_{2}$Se$_{3}$ which makes accessing
the TSS via electrostatic gating easier~\citep{RN1234,yoshimi2015quantum}.
Gate-dependent PNR measurements performed by Li \emph{et al.}~\citep{RN688}
showed that the magnetic proximity effect in EuS--(Bi$_{0.2}$Sb$_{0.8}$)$_{2}$Te$_{3}$
is maximized at the Dirac point, \emph{i.e} at low carrier density. The authors attributed this behavior to paramagnetic nature of TSS and diamagnetic nature of bulk electrons, and their respective magnetic screening behavior. Surprisingly, EuS--(Bi$_{0.2}$Sb$_{0.8}$)$_{2}$Te$_{3}$ heterostructures have not shown any unambiguous electrical transport
signature of a magnetized system such as weak localization or AHE~\citep{wei2013exchange}.
However, Yang \emph{et al}.~\citep{RN692} reported concurrent
resistance transitions and magnetic anomalies in such heterostructures
at temperatures well above $T_{c}$ of EuS, suggesting a connection between magnetic order and transport in the topological layer. Though a crossover between positive
and negative magnetoresistance was reported in ultrathin films of (Bi$_{0.05}$Sb$_{0.95}$)$_{2}$Te$_{3}$,
tunnel coupling between surfaces is expected to gap the surface states~\citep{RN692}, and indeed the conductance was below the Mott-Ioffe-Regel
limit $e^{2}/h$. Hence, the crossover might have been caused by strong localization, rather than a topological phase transition. Understanding the true nature of room temperature
interface-magnetization is required to harness the full potential
of these materials in the future.

\paragraph{\textbf{Rare-earth garnets:}} Rare earth garnets such as YIG (Fe$_{5}$Y$_{3}$O$_{12}$),
TIG (Tm$_{3}$Fe$_{5}$O$_{12}$) are another alternative family of FMIs for FMI--TI
heterostructures as these are ferrimagnetic insulators with very high $T_{c}$
($550$~K)~\citep{neel1964chapter}. The first attempt at synthesizing
heterostructures of YIG and Bi$_{2}$Se$_{3}$~\citep{RN973} showed
(i) a clear electrical transport signature of magnetic proximity coupling, and
(ii) a Magneto-optical Kerr effect (MOKE) signature of magnetic anisotropy
up to $130$~K with spin orientation antiparallel to the magnetization
of YIG. Although bulk YIG is known to have in-plane anisotropy,
the temperature-dependent lattice constant mismatch between YIG and substrate GGG (Gd$_{3}$Ga$_{5}$O$_{12}$)
used in this case gave rise to an out-of-plane component of magnetization,
which is ideal to gap the TI surface state. Heterostructures
of YIG and (Bi$_{x}$Sb$_{1-x}$)$_{2}$Te$_{3}$ have demonstrated clear
signatures of magnetic proximity effect at different composition (by
varying $x$) and carrier density~\citep{RN1005,RN1004} through AHE and suppression of WAL.

\selectlanguage{english}%
TIG films grown on (111)-oriented substituted gadolinium gallium garnet (SGGG) substrate by pulsed laser deposition are known to have perpendicular magnetic anisotropy caused by strain~\citep{tang2016anomalous}. In a SGGG--TIG--TI heterostructure (Figure~\ref{fig:Polarized-neutron-reflectivity}d), the out-of-plane
magnetization of the TIG induced robust AHE in (Bi$_{x}$Sb$_{1-x}$)$_{2}$Te$_{3}$
($5$~QL) with $T_{c}>400$~K~\citep{Tange1700307}. Figure~\ref{fig:Polarized-neutron-reflectivity}e)
and f) shows AHE measurements in TIG--(Bi$_{x}$Sb$_{1-x}$)$_{2}$Te$_{3}$ heterostructure for $x=0.2$ and $x=0.3$. To check that the AHE loops are not an artifact of a spin Hall effect
(as has been seen in the TIG--Pt system~\citep{tang2016anomalous}),
point-contact Andreev reflection spectroscopy was employed to detect
bottom-surface spin-polarization, revealing proximity-induced ferromagnetism
in the TI surface electronic states. This extraordinary result shows that FMI--TI heterostructures are capable of breaking time-reversal symmetry in a TI at room
temperature, far above the temperatures achieved to date in magnetically doped TIs and intrinsic MTIs. 

However, while the above and other measurements~\citep{RN664} demonstrate a magnetic proximity effect experimentally, they do not provide direct evidence of exchange gap opening.  More recently magnetoconductance measurements were utilized to detect the presence of an exchange-gapped Dirac cone in both Bi$_{2}$Se$_{3}$--YIG and Bi$_{2}$Se$_{3}$--TIG
heterostructures~\citep{RN939} (further details of magnetoconductance measurements in these types of systems will be discussed in the next subsection). Apart from
the promising systems discussed above, other insulating ferromagnetic oxides like LCO (LaCoO$_{3}$) and
CFO (CoFe$_{2}$O$_{4}$) have been explored for FMI--TI heterostructures, and preliminary
data in these materials show evidence of proximity effect and exchange gap~\citep{RN592,RN941}. 

\subsubsection{Dilute ferromagnetic semiconductor (DFMS)}
\label{subsection:Dilute ferromagnetic semiconductor}

Dilute ferromagnetic semiconductors (DFMS) are semiconductors dilutely doped
by transition metal atoms such as Cr, Fe, or Mn. These materials have
dual advantages of semiconductors and magnetic materials (MMs); \emph{i.e.} both
electronic properties and spin-orientation can be manipulated in these materials~\citep{macdonald2005ferromagnetic,furdyna1988diluted}.
Moreover, magnetic, and electrical properties in these materials are
coupled, hence the magnetic state in these materials can be controlled by electric field~\citep{macdonald2005ferromagnetic,ohno1998making}.
These materials also may be easier to synthesize compared to intrinsic
ferromagnetic insulators (FMIs) because their parent semiconductors,
such as GaAs, InAs, ZnO, CdTe, ZnSe are well-understood, well-characterized
and have well-established growth recipes. These materials have presented
themselves to be extremely capable for semiconductor spintronics applications~\citep{ohno2004ferromagnetic}, and in some cases $T_c$ as high as 300 K and 340 K have been achieved~\citep{tu2016high}, enabling the possibility of room temperature applications.

Ga$_{1-x}$Mn$_{x}$As, a III-V magnetic semiconductor, is
one of the most well-studied dilute magnetic semiconductor~\citep{macdonald2005ferromagnetic}.
In this material, 3$d$ transition metal Mn is substituted for Ga
in the GaAs lattice. As GaAs is already being used
in commercial electronic devices, Ga$_{1-x}$Mn$_{x}$As has a potential
to be integrated in these devices for practical applications. The Mn
atoms in this lattice have a dual contribution: (1) the half-filled $d$-shell
of Mn$^{2+}$ provides local magnetic moment with spin $S=\frac{5}{2}$, while
(ii) Mn impurities form an acceptor level, which provides holes that
mediate a ferromagnetic interaction between the Mn$^{2+}$ local moments.
The $T_c$ of this material can be increased by increasing Mn concentration
and $T_{c}$ well above 150~K has been reported so far~\citep{macdonald2005ferromagnetic}. 
\begin{figure}[!]
\includegraphics[width=\columnwidth]{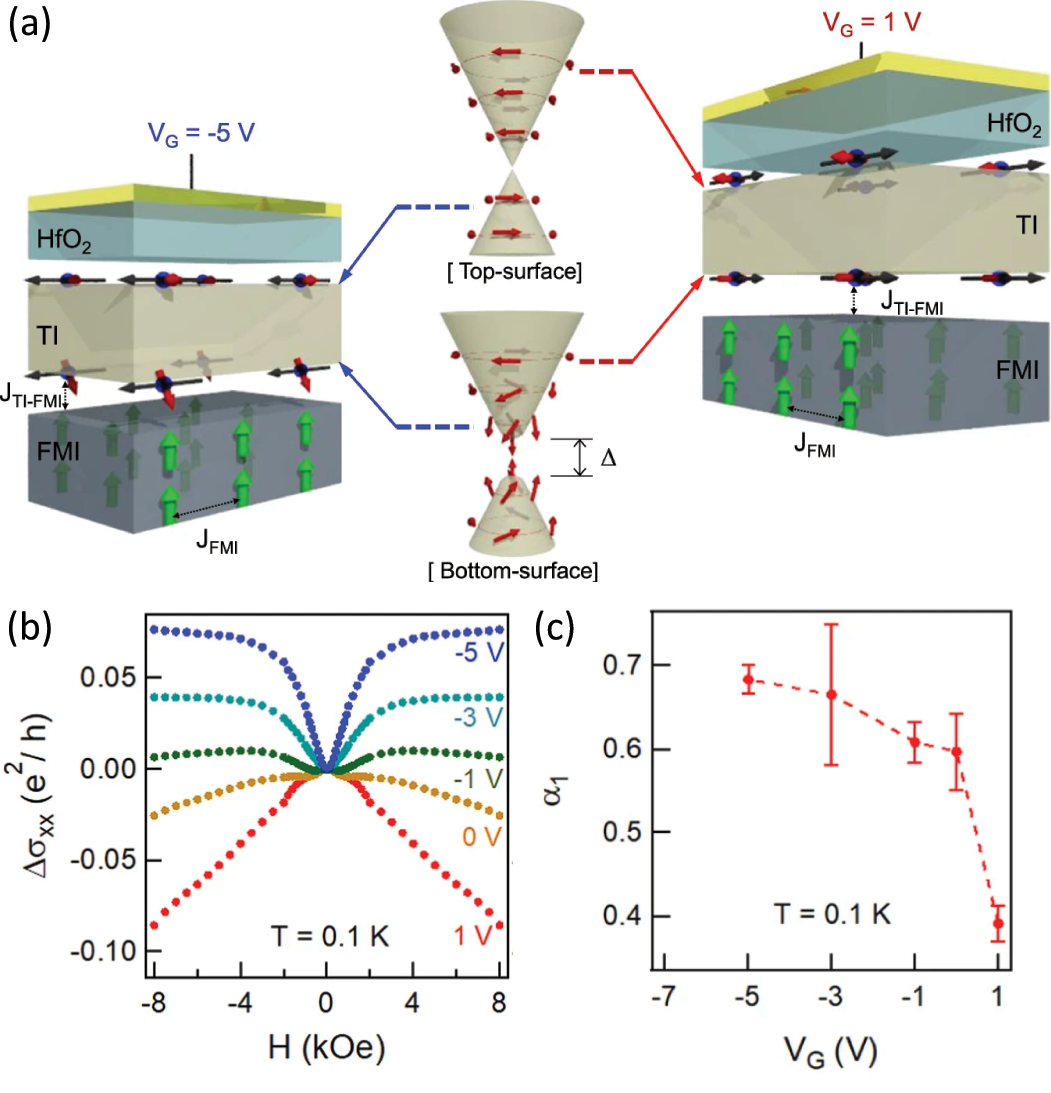}

\caption{Gate-dependent magnetotransport measurements in dilute ferromagnetic semiconductor (DFMS)--topological insulator (TI) heterostructure (Ga,Mn)As--(Bi,Sb)$_{2}$(Te,Se)$_{3}$. a) Band structure and spin textures in TI surfaces. Illustration of a gapless Dirac surface state
for the top surface, a magnetically gapped surface state with hedgehog
spin texture for the bottom surface, and top-gated (Ga,Mn)As--(Bi,Sb)$_{2}$(Te,Se)$_{3}$ heterostructure devices with $V_{G}=-5$~V and $V_{G}=+1$~V. Green
arrows represent the magnetization of the (Ga,Mn)As layer and gray
and red arrows in the TI layer represent the directions of electron
propagation and its spin, respectively. $\textrm{J}_{\textrm{FMI}}$
is the exchange-coupling constant between Mn moments in (Ga,Mn)As
while $\textrm{J}{}_{\textrm{TI-FMI}}$ is the weaker exchange-coupling
constant between electrons in the TI bottom surface and Mn moments
in (Ga,Mn)As. b) Magnetoconductance $\Delta\mathrm{\sigma_{xx}}$(H) at different applied gate voltage ($V_{G}$) as indicated on the plot
at $T=0.1$~K . A crossover from negative magnetoconductance (WAL) to positive magnetoconductance
(WL) is observed as $V_{G}$ decreases from $1$ to $-5$~V. c) Gate-voltage dependence of the
prefactor $\alpha_{1}$ for the bottom surface obtained by fitting the
MC in b) at $T = 0.1$~K to \cref{eq:MC-1}. Adapted
from Ref.~\citep{RN647} under Creative Commons licenseAttribution 4.0 International License, (\url{https://creativecommons.org/licenses/by/4.0/legalcode}).\label{fig:(a)-Lateral-electrical}}
\end{figure}
The magneto-transport properties of (Bi,Sb)$_{2}$(Te,Se)$_{3}$--Ga$_{1-x}$Mn$_{x}$As
were studied in depth by Lee \emph{et al.}~\citep{RN647} as a function
of gate voltage and temperature. Figure~\ref{fig:(a)-Lateral-electrical}a) shows the schematic of a typical device investigated in this study. The doping level of Mn, the substrate,
and the growth method were appropriately chosen in order to synthesize
a highly-resistive Ga$_{1-x}$Mn$_{x}$As with perpendicular magnetic
anisotropy. A HfO$_{2}$ top gate was used
to control the chemical potential of (Bi,Sb)$_{2}$(Te,Se)$_{3}$. This top gate successfully
tuned the chemical potential at both top and bottom surface, albeit
with a higher degree of control at the top surface compared to the
bottom. However, the proximity-induced
gap-opening is expected to occur only at the bottom surface, while the
top surface remains gapless. Figure~\ref{fig:(a)-Lateral-electrical}a) additionally illustrates the band structures of both surfaces, and  the effect of top and gate on the chemical potential of ungapped top surface and gapped bottom surface. 

Figure~\ref{fig:(a)-Lateral-electrical}b) shows magnetoconductance measurements in this system at varying top gate voltage. The magnetoconductance data shows a crossover from negative to positive magnetoconductance as the gate voltage ($V_G$) is varied from 1~V to -5~V, \textit{i.e.} the chemical potential is tuned from conduction band towards valence band. Adapting the Hikami-Larkin-Nagaoka (HLN)
theory~\citep{hikami1980spin}, this data is fitted with the phenomenological equation
\begin{equation}
\Delta\sigma(H)=\underset{}{\sum_{\mathop{i=0,1}}\frac{\alpha_{i}e^{2}}{2\pi^{2}\hbar}\left[\psi\left(\frac{1}{2}+\frac{\hbar c}{4el_{\phi,i}^{2}H}\right)-\ln\left(\frac{\hbar c}{4el_{\phi,i}^{2}H}\right)\right]}\label{eq:MC-1} 
\end{equation}

where, $e$ is electronic charge, $\hbar$ is reduced Planck's constant,
$\psi$ is the digamma function, and $l_{\phi,i}$ and
$\alpha_{i}$ are the dephasing lengths and prefactors respectively,
corresponding to the topmost gapless surface state ($i=0$) and the
bottom gapped surface state ($i=1$). As the top gapless surface
state of TI is a metallic diffusive system with spin-orbit coupling
and $\pi$ Berry phase, according to HLN theory such a system is expected
to have $\alpha_{0}=-\frac{1}{2}$, indicating perfect weak antilocalization, and negative magnetoconductance behavior.
The bottom surface ($i=1$) is modeled with a variable $\alpha_{1}$,
which phenomenologically is expected to vary between the two well understood limits of $\alpha_{1}=-\frac{1}{2}$ for a
gapless state, and $\alpha_{1}=1$, for a conventional semiconductor
with no spin-orbit coupling and $0$ Berry phase, with weak localization and positive magnetoconductance.

Figure~\ref{fig:(a)-Lateral-electrical}c) shows the fitted value of $\alpha_{1}$ as a function of $V_{G}$. The increase of $\alpha_{1}$ with decreasing $V_{G}$ can be explained by the change of Berry phase with $V_{G}$. $V_{G}$ controls chemical potential $E_{F}$, which in turn tunes the Berry phase $\left(\pi\left(1-\frac{\Delta}{2E_{F}}\right)\right)$, where ${\Delta}$ is the exchange gap, and this
results in a gradual WAL-WL crossover. Correspondingly, the anomalous Hall component of $R_{xy}$ increases with decreasing gate voltage (not shown here), also suggesting a gapped surface state. Additionally, the temperature dependence
of both magnetoconductance and Hall data showed a gradual reduction
with increasing temperature, reflecting the temperature dependence
of the interfacial exchange coupling with the adjacent (Ga,Mn)As layer.
This study not only introduced (Ga,Mn)As, a well-optimized magnetic
semiconductor, as a good candidate as FMI in FMI--TI heterostructure,
but also demonstrated a systematic way of conducting and analyzing
magnetotransport measurements in gate-tunable FMI--TI heterostructures in general.

Bac \emph{et al.}~\citep{RN586} however demonstrated that one needs
to be careful while analyzing the transport data of such heterostructures, as
the TI can modify the electrical and magnetic behavior of Ga$_{1-x}$Mn$_{x}$As.
However, it is important to note, the Ga$_{1-x}$Mn$_{x}$As used
in the latter study was considerably more doped compared to the former study,
and the TI used in this case was Bi$_{2}$Se$_{3}$,
which is known to be naturally doped with the chemical potential in the bulk conduction band~\citep{analytis2010bulk}.
Thus the modification of the properties of Ga$_{1-x}$Mn$_{x}$As observed by Bac \emph{et al.} might
have resulted from diffusion of highly dense Mn atoms, or charge transfer
from highly doped Bi$_{2}$Se$_{3}$ to Ga$_{1-x}$Mn$_{x}$As. Nonetheless, III-V based dilute magnetic semiconductors are an extremely important system to be considered due to their versatility and well-understood characteristics.

\subsubsection{Van der Waals layered ferromagnetic materials}
\label{subsection:van der Waals layered ferromagnetic materials}

Recently discovered van der Waals magnetic materials (MMs)~\citep{burch2018magnetism,ningrum2020recent},
\emph{i.e.}, layered materials which contain 2D sheets of magnetic
materials stacked with van der Waals forces, have brought a paradigm
shift in heterostructures of ferromagnetic materials, opening the possibility of van der Waals heterostructures~\citep{geim2013van,novoselov20162d,RN1033} incorporating magnetism. According to theoretical calculations, ferromagnetism can be realized in several types of layered transition metal compounds, including transition metal trichalcogenides (CrGeTe$_{3}$, CrSiTe$_{3}$ etc.), transition metal dichalcogenides (MnS$_{2}$, VSe$_{2}$ etc.), trihalides (CrI$_{3}$, MnCl$_{3}$ etc.), dihallides (FeCl$_{2}$, NiI$_{2}$ etc.), and other materials, such as, MnO$_{2}$ and FeC$_{2}$~\citep{RN1036}. These materials are also predicted to display a wide range of electrical properties, from metallic, half-metallic, to semiconducting, and insulating. Some of these materials are predicted to show extremely high \emph{T$_{c}$} (1877~K in half-metallic Mn$_{2}$NF$_{2}$), and intrinsic magnetism down to monolayer thickness, albeit, in some cases, with a lower \emph{$T_{c}$}. These materials are discussed in details in several recent review articles~\citep{RN1036,RN1032, ningrum2020recent,RN1031,RN1033}. 

Of particular interest for MM--TI heterostructures are the  semiconducting and insulating van der Waals ferromagnets. Some of these are already experimentally realized, such as Cr$_{2}$Ge$_{2}$Te$_{6}$~\citep{gong2017discovery}, CrI$_{3}$~\citep{huang2017layer}, and especially VS$_{2}$ which has showed room temperature ferromagnetism at the monolayer limit~\citep{doi:10.1002/adma.201700715}. For a complete and updated list of these materials see ref.~\citep{RN1032}. In addition to these, intrinsic magnetic
TIs, such as, CrBi$_{2}$Se$_{4}$~\citep{RN612}, and superlattice structures, such as MnBi$_{8}$Te$_{13}$~\citep{Hueaba4275}, as well as dilute magnetic doped TI have both surface as well as bulk band gaps, and some of these are conventional insulators in the monolayer limit. Hence, these materials 
can also be used as the ferromagnetic insulator (FMI) in such
heterostructures~\citep{RN648, Mogieaao1669}.

Van der Waals MMs have several advantages for MM--TI heterostructures: 

(i) Due to their layered structure, and lattice matching in some cases with
TI materials which are layered themselves, ultraflat defectless interfaces
of these materials can be obtained by simple growth procedures. Due
to the absence of dangling bonds, it is also possible to implement
van der Waals epitaxy for these heterostructures, which does not require lattice matching.

(ii) Magnetic atoms of these materials are usually not located in
the outer-most layer which suppresses diffusion of such atoms inside the
neighboring TI layers.

(iii) Magnetic and electrical properties in these materials can be
altered through chemical doping, layer number change, strain or proximity
effects.

\begin{figure*}{t}
\includegraphics[width=\textwidth]{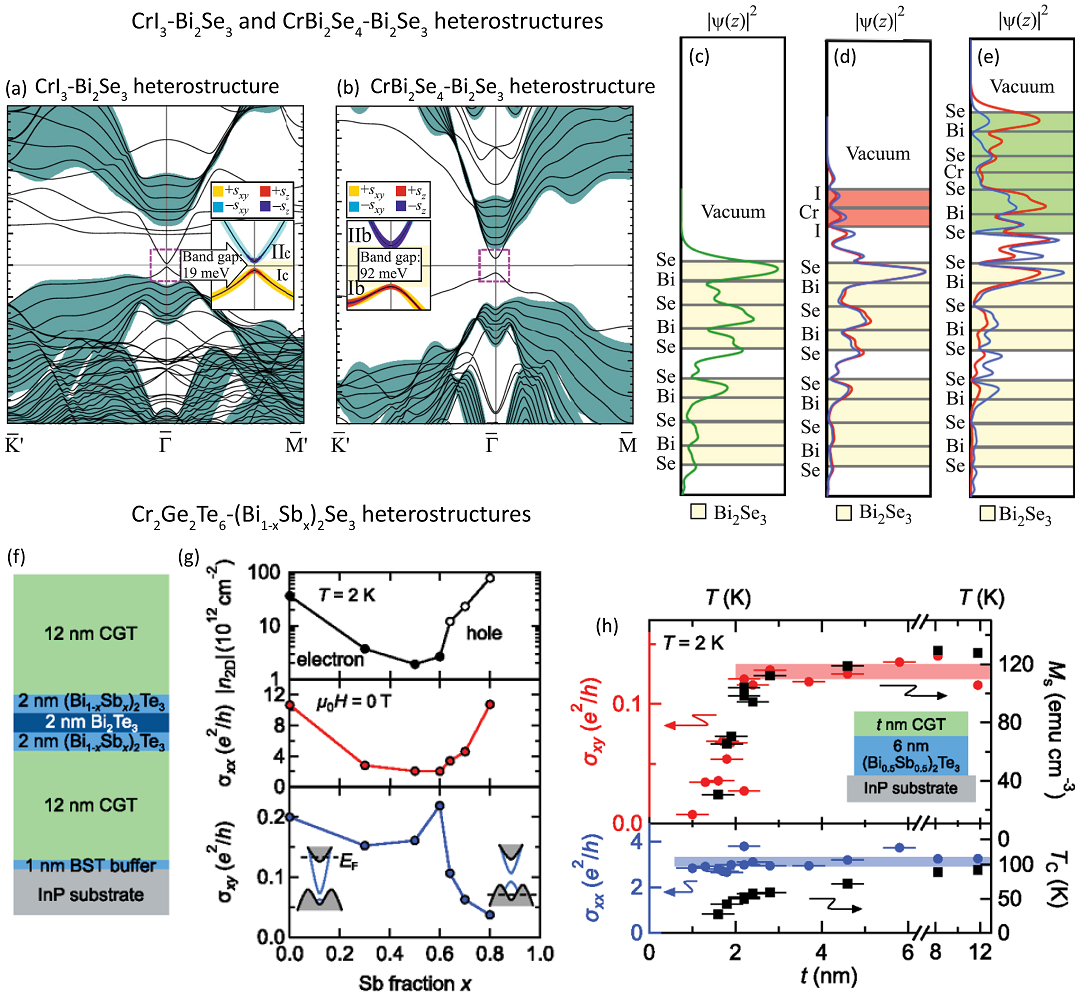}

\caption{Van der Waals ferromagnetic insulator (vdWFMI)--Topological insulator (TI)  heterostructure. a-c) Calculation of band structure and wave-function at the interface of Cr-containing ferromagnetic material-TI heterostructure. Band structure of the (a) CrI$_{3}$--Bi$_{2}$Se$_{3}$ and (b) CrBi$_{2}$Se$_{4}$--Bi$_{2}$Se$_{3}$
heterostructures. Black lines show the band structure of the considered
films. The projections of bulk Bi$_{2}$Se$_{3}$ states on the two-dimensional
Brillouin zone of films are shown in dark green. The inset shows the
spin texture of bands near the Fermi level, where the thickness of
a color line is proportional to the corresponding spin projection.
c-e) Square of the absolute value of the wavefunction $\left|\Psi(z)\right|^{2}=\int\int\left|\Psi\right|^{2}\left(x,y,z\right)dxdx$
at the $\Gamma$ point integrated in the $x-y$ plane for (c) Bi$_{2}$Se$_{3}$
compounds, (d) CrI$_{3}$--Bi$_{2}$Se$_{3}$, and (e) CrBi$_{2}$Se$_{4}$--Bi$_{2}$Se$_{3}$
heterostructures. The green line in panel (c) is the charge density
of the topological state at the Dirac point. Red and blue lines in
panels (d) and (e) are calculated for states Ic and Ib of the top of
the valence band and for states IIc and IIb of the conduction band. f-h) Anomalous Hall effect in trilayer heterostructure of Cr$_{2}$Ge$_{2}$Te$_{6}$ (CGT)--TI--CGT. f) Schematic layout of the heterostructure consisting of 12~nm CGT--2~nm (Bi$_{1-x}$Sb$_{x}$)$_{2}$Te$_{3}$--2~nm Bi$_{2}$Te$_{3}$--2~nm (Bi$_{1-x}$Sb$_{x}$)$_{2}$Te$_{3}$--12~nm CGT
heterostructure.
g) Sb fraction ($x$) dependence of the sheet carrier density ($\left|n_{2D}\right|$)
(top), the longitudinal sheet conductivity ($\sigma_{xx}$) (middle),
and the Hall conductivity ($\sigma_{xy}$) (bottom) at $2$~K. Insets shows simplified schematics of band structures representing
the different Fermi energies; the blue lines represent the dispersion
of the surface state. h) CGT thickness $t$ dependence of $\sigma_{xy}$
(red closed circles, left red ordinate) and the spontaneous magnetization
$M_{s}$ (black closed squares, right black coordinate) at $2$~K under
zero magnetic field (top) and $\sigma_{xx}$ at $2$~K (blued closed
circles, left blue ordinate) and $T_{c}$ (black closed squares, right
black ordinate) (bottom). Inset: Schematic layout of (Bi$_{0.5}$Sb$_{0.5}$)$_{2}$Te$_{3}$
(6~nm)--CGT ($t$~nm) bilayer structure. a-e) Reprinted by permission RightsLink Printable License: Springer Nature, JETP Letters , Cr-Containing Ferromagnetic Film-Topological Insulator Heterostructures as Promising Materials for the Quantum Anomalous Hall Effect~\citep{RN612}, E. K. Petrov, I. V. Silkin, T. V. Menshchikova and E. V. Chulkov, \copyright Pleiades Publishing, Inc., (2019). Reprinted f-h) with permission from~\citep{RN637} Masataka Mogi, Taro Nakajima, Victor Ukleev, Atsushi Tsukazaki, Ryutaro Yoshimi, Minoru Kawamura, Kei S. Takahashi, Takayasu Hanashima, Kazuhisa Kakurai, Taka-hisa Arima, Masashi Kawasaki, and Yoshinori Tokura, Physical Review Letters 123, 016804 (2019) Copyright (2019) by the American Physical Society \url{(DOI:10.1103/PhysRevLett.123.016804)}.\label{fig:Band-structure-of} }

\end{figure*}

Theoretical studies have shown that, despite the lack of strong bonding across the MM--TI van der Waals interface, strong magnetic proximity effects are possible. Petrov \emph{et al.} studied heterostructures of both van der Waals
ferromagnet monolayer-CrI$_{3}$, and intrinsic MTI
CrBi$_{2}$Se$_{4}$ with Bi$_{2}$Se$_{3}$~\citep{RN612} through
ab-initio calculations. Figure~\ref{fig:Band-structure-of}a) and b) show the band structures at the interfaces. The surface
states of Bi$_{2}$Se$_{3}$ exhibit bandgaps of 19~meV and 92~meV respectively for proximity with CrI$_{3}$ and CrBi$_{2}$Se$_{4}$. The bandgap opening is attributed
to redistribution of reconstructed TSS into
the neighboring FMIs. This is illustrated in Figure~\ref{fig:Band-structure-of}c-e).
The probability densities of the gapped TI surface states have a larger spatial
extension into CrI$_{3}$ compared to that of gapless surface states
into vacuum. Petrov \emph{et al.} found an even larger extension in the CrBi$_{2}$Se$_{4}$--Bi$_{2}$Se$_{3}$
heterostructure (Figure~\ref{fig:Band-structure-of}e) , where a significant redistribution of the charge
density is observed. This phenomenon, resulting from similar lattice
and electronic structure of the two materials, known as magnetic extension,
gives rise to higher exchange gap for heterostructures involving FMI with similar crystal structure and chemical composition~\citep{otrokov2017highly,RN686}. In another study, theoretical
calculations showed even monolayer CrI$_{3}$ can create a sizable
exchange gap at the charge neutrality point of TI surface~\citep{Houeaaw1874}.

The first successful growth of a Cr$_{2}$Ge$_{2}$Te$_{6}$ (CGT)--Bi$_{2}$Se$_{3}$-family heterostructure
was achieved by Mogi \emph{et al.}~\citep{RN663} through MBE. These heterostructures
develop a large remanent magnetization with perpendicular anisotropy
and $T_{c}=80$~K. Mogi \emph{et al.} have further probed CGT--TI--CGT
heterostructures with PNR measurements and electrical transport measurements~\citep{RN637}.
Figure~\ref{fig:Band-structure-of}f) shows a schematic of the heterostructure studied using electrical measurements by  Mogi \emph{et al.}~\citep{RN637}; the TI layer consists
of 2~nm (Bi$_{1-x}$Sb$_{x}$)$_{2}$Te$_{3}$--2~nm Bi$_{2}$Te$_{3}$--2~nm
(Bi$_{1-x}$Sb$_{x}$)$_{2}$Te$_{3}$. The number density and chemical
potential of this TI layer was varied by changing $x$. Figure~\ref{fig:Band-structure-of}g) shows the number density $n_{2D}$, sheet-conductance ($\sigma_{xx}=2$~e$^{2}$/h), and zero-field Hall conductivity ($\sigma_{xy}=0.2$~e$^{2}$/h) as a function of $x$. The anomalous
Hall effect which appears in all samples with $x=0.3-0.64$ shows
a maximum of Hall conductivity ($\sigma_{xy}=0.2$~e$^{2}$/h), near-minimum
sheet-conductance ($\sigma_{xx}=2$~e$^{2}$/h), and a record Hall
angle ($\theta_{H}=\tan^{-1}\left(\sigma_{xy}/\sigma_{xx}\right)$)
for FMI--TI heterostructure at $x=0.6$ which
has very low number density $n_{2D}=10^{12}$~cm$^{-2}$.
This indicates that the magnetic proximity effect has opened an exchange gap at both
the surface states of the TI. When the Fermi energy is tuned near the
band edge, the AHE is increased due to the modification of the Berry
curvature $\left(\pi\left(1-\frac{\Delta}{2E_{F}}\right)\right)$. Although electronic transport measurements show convincing
evidence for an exchange gap, according to PNR measurements, the induced
magnetization at the TI-interface is very small. In order to ensure
this is a genuine proximity effect and not caused by any artifacts
like Cr diffusion, Mogi \emph{et al.} also measured magnetization and AHE at CGT/TI heterostructures with varying thickness ($t$) of CGT. Figure~\ref{fig:Band-structure-of}h) demonstrates a saturation of $\sigma_{xy}$,
$T_{c}$ and $M_{s}$ above $t=2$~nm, and a systematic
reduction with $t$ below $t<2$~nm due to finite size effect of
the 2D ferromagnetic CGT layer. This provides convincing evidence that Cr-diffusion is
not responsible for the exchange gap formation in this sample, and
thus the exchange gap formation is attributable to magnetic
extension of TI surface states in CGT. 

The field of layered van der Waals ferromagnetic insulators (vdWFMIs) is in its infancy compared to the well-established fields of lanthanide-based, garnet-based, transition metal oxide-based ferromagnetic semiconductors or dilute magnetic insulator. Yet, vdW FMI--TI--FMI heterostructures are already showing remarkable promise. Still the versatility of van der Waals heterostructures has only just begun to be explored in the context of MM--TI heterostructures, compared to other materials (such as graphene and 2D semiconductors). Later in this review, we will discuss how new fabrication techniques such as mechanical van der Waals heterostructure fabrication, or wet transfer techniques can help propel this field forward (section~\ref{section:Conclusion and Outlook}).

\subsection{Antiferromagnetic proximity effect}
\label{section:Anti-ferromagnetic proximity effect}

\begin{figure*}
\includegraphics[width=\textwidth]{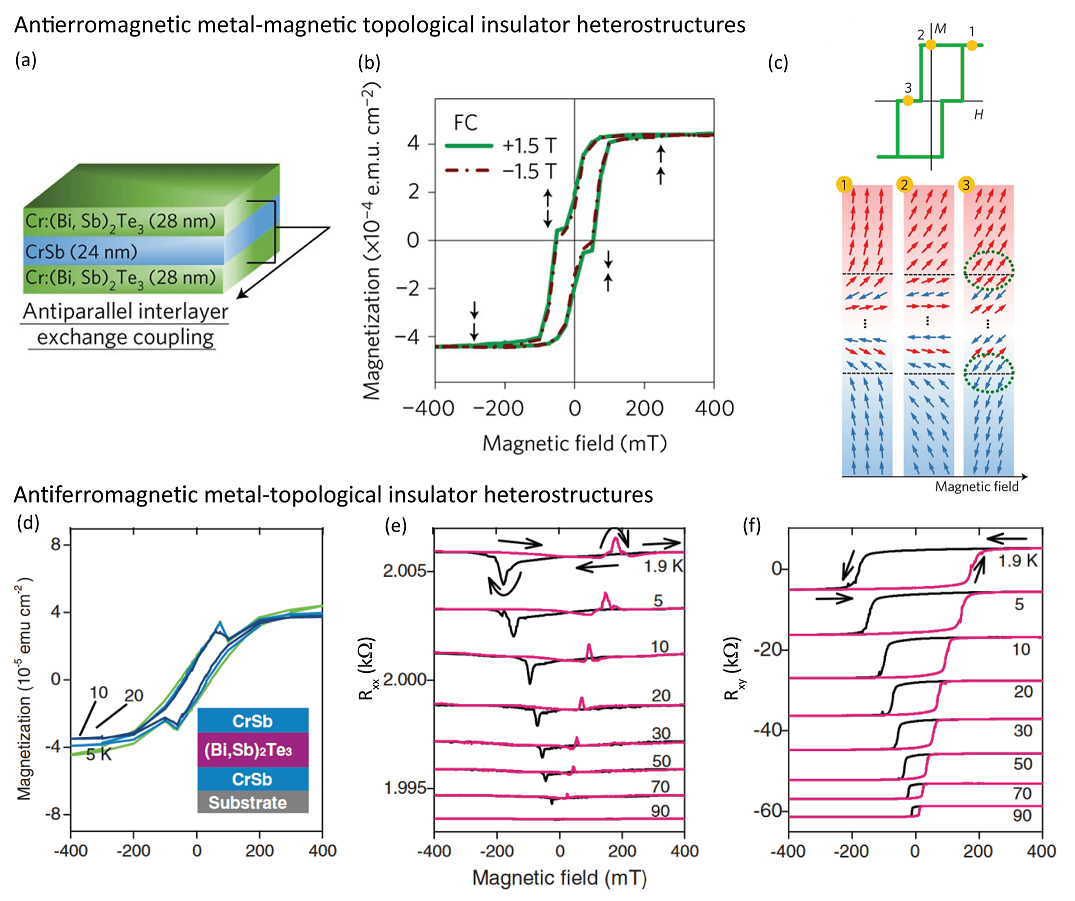}

\caption{Antiferromagnetic proximity effect. a-c) Proximity coupling in antiferromagnetic metal (AFMM)--magnetic topological insulator (MTI) heterostructure. a) Schematic of magnetic topological insulator MTI--AFMM--MTI trilayer Cr-doped (Bi,Sb)$_{2}$Te$_{3}$--CrSb--Cr-doped (Bi,Sb)$_{2}$Te$_{3}$. b) Magnetization vs. magnetic field $M(H)$ loop of the MTI--AFMM--MTI trilayer. c) Schematic of the symmetrical magnetization loop with a double-switching feature as observed in (b), which indicates a novel antiparallel effective long-range exchange coupling mediated by the interactions between the MTI surface spins and AFM spins. The relative orientations of the atomic moments of the effective long-range exchange coupling in the MTI--AFMM--MTI trilayer are shown for three different positions along the $M(H)$ loop as indicated by numbers. d-f) Proximity coupling in an AFMM--topological insulator (TI) heterostructure. d) Temperature-dependent
$M(H)$ loops in AFMM--TI--AFMM heterostructure CrSb--(Bi,Sb)$_{2}$Te$_{3}$--CrSb. Magnetic field dependence of e) the longitudinal ($R_{xx}$)
resistance, and (f) the Hall ($R_{xy}$) resistance in this AFMM--TI--AFMM
trilayer, showing an antisymmetric magnetoresistance (MR) behavior and an anomalous Hall effect (AHE). Temperatures indicated on each plot next to the corresponding curve. 
Traces are offset for clarity, except for the traces at 1.9 K. a-c) Reprinted by permission from RightsLink Printable Licenses: Springer Nature Nature Materials Tailoring exchange couplings in magnetic topological-insulator/antiferromagnetic heterostructures~\citep{RN611}. Qing Lin He, Xufeng Kou, Alexander J. Grutter, Gen Yin, Lei Pan, Xiaoyu Che, Yuxiang Liu, Tianxiao Nie, Bin Zhang, Steven M. Disseler, Brian J. Kirby, William Ratcliff II, Qiming Shao, Koichi Murata, Xiaodan Zhu, Guoqiang Yu, Yabin Fan, Mohammad Montazeri, Xiaodong Han, Julie A. Borchers and Kang L. Wang,  \copyright~Macmillan Publishers Limited (2016). d-f) Reprinted with permission from Ref~\citep{RN694} Qing Lin He \emph{et al.}, Physical Review Letters, 121, 096802 (2018) Copyright 2018 by the American Physical Society \url{(DOI: 10.1103/PhysRevLett.121.096802)}}.{\label{fig:AFM}}
\end{figure*}

So far we have focused on the proximity effect of ferromagnets (FMs) and TIs. However, as discussed before in section~\ref{subsubsection:Direct exchange coupling}, the interfacial nature of proximity-induced magnetism enables antiferromagnetism as well to break the TRS and generate massive Dirac surface states in a TI, which in turn can lead to exotic topological states such as QAHI and axion insulator states~\citep{RN904,yu2010quantized,chang2013experimental,RN622}. Additionally, there are many more antiferromagnetic insulators (AFMIs) than ferromagnetic insulators (FMIs) known, and the N\'{e}el temperatures of AFMIs are often higher than the Curie temperatures of FMIs~\citep{jungwirth2016antiferromagnetic}. Also, AFMIs have a very small net magnetization, avoiding strong stray fields which may affect the TI layer. Therefore, magnetic order remains robust when the external magnetic field is applied.

However, even though AFMIs have several clear advantages over FMIs, AFMI-based heterostructures have received relatively less attenion. Lau and Qi performed DFT-calculations to show that MnSe, which hosts ferromagnetically-ordered spins in the [111] plane, is an even better candidate to induce magnetism in TI, compared to the FMI EuS~\citep{RN916}. Following that, Matetskiy \emph{et al.} experimentally showed a gap-opening of 90~meV at the interface of MnSe--Bi$_{2}$Se$_{3}$ through ARPES measurement~\citep{matetskiy2015direct}. However, magnetotransport and magnetometry measurements in these heterostructures are still lacking.

Surprisingly, tight-binding calculations by Oroszl{\'a}ny and Cortijo showed that even anantiferromagnetic interface with staggered magnetization can lead to proximity-induced gap opening in TI surface states~\citep{oroszlany2012gap}, making AFMI--TI heterostructures even more promising. Magnetotransport in such a heterostructure (NiO--Bi$_{2}$Se$_{3}$) showed suppressed WAL, and even WL at a low magnetic field in a sample with ultrathin (5~nm) TI~\citep{RN589}. Whereas these results are encouraging, in order to unambiguously detect the proximity effect, further experimental investigations are necessary.

Magnetic proximity coupling has been successfully achieved in antiferromagnetic metal--topological insulator (AFMM--TI) heterostructures involving AFMM CrSb~\citep{suzuoka1957electrical,allen1976electronic,RN611, RN694}. Figure~\ref{fig:AFM}a) shows the schematic of magnetic topological insulator--antiferromagnetic metal--magnetic topological insulator (MTI--AFMM--MTI) trilayer heterostructure exhibiting antiferromagnetic order in CrSb and ferromagentic order in Cr-doped (Bi,Sb)$_2$Te$_3$. He \emph{et al.} have shown this trilayer heterostructure exhibits a unique $M(H)$ loop (Figure~\ref{fig:AFM}b) showing double switching, which results from an effective long-range exchange coupling between MTI layers mediated by the antiferromagnetic layer~\citep{RN611}. Figure~\ref{fig:AFM}b) shows the $M(H)$ loop, which exhibits an additional step with near-zero magnetization. Figure~\ref{fig:AFM}c) show the $M(H)$ loop schematically along with the spin configurations for different points along the loop. The near-zero magnetization step (point 3 in Figure~\ref{fig:AFM}c)) results from an antiparallel configuration of the magnetizations in the two Cr-doped (Bi, Sb)$_2$Te$_3$ layers. This state is normally unfavorable energetically, and the authors did not observe this state in a similar heterostructure of two Cr-doped (Bi,Sb)$_2$Te$_3$ with an intervening non-magnetic (Bi,Sb)$_2$Te$_3$ layer. However it can be stabilized by local ferrromagnetic exchange coupling of each of the Cr-doped (Bi,Sb)$_2$Te$_3$ layers to different spin sublattices in the CrSb antiferromagnet (AFM) at the interfaces, enabled by spin canting near the interfaces, which can reduce the exchange energy and make the configuration energetically favorable. 

This study indicates that the N\'{e}el temperature might be increased by manipulating interfacial magnetic interactions, and demonstrates that proximity induced surface state magnetization can occur at the interface between AFMs and MTIs~\citep{RN611}. In principle, this is due to short-range interfacial exchange coupling to an uncompensated AFM plane that can locally magnetise the surface of the TI~\citep{snow1953magnetic}. However, the heterostructure effectively engineers a long-range exchange coupling between the MTI layers using the AFM layer with N\'{e}el temperature up to $\sim{700}$~K. Heterostructure of AFM and MTI may also lead to additional manipulation of the quantum anomalous Hall insulator (QAHI) state through exchange bias effect~\citep{pan2020observation}.

In a more recent study, He \emph{et al.} has shown that the top and the bottom TI surface state can be magnetized by antiferromagnetic thin films through interfacial coupling~\citep{RN694}. Figure~\ref{fig:AFM}d) shows the $M(H)$ behavior of an AFMM--TI--AFMM heterostructure of CrSb--(Bi, Sb)$_2$Te$_3$--CrSb, shown schematically in the inset. Surprisingly, the $M(H)$ shows a clear hysteresis typical of a ferromagnet (hystersis was not seen in a bare CrSb film or CrSb--(Bi, Sb)$_2$Te$_3$ heterostructure). Figure~\ref{fig:AFM}e) shows that, antisymmetric magnetoresistance spikes develop around the field at which the magnetization reversal is observed in $M(H)$, while Figure~\ref{fig:AFM}f) shows the the AHE exhibits a typical ferromagnetic response, consistent with the $M(H)$. 
The authors interpret the behavior as the result of a proximity-induced ferromagnetic order in the top and bottom TI surfaces states, which leads to observation of AHE at temperatures up to 90 K. A sequential reversal of the magnetizations of the two surfaces results in intermediate magnetization arrangements giving rise to the resistance spikes in Figure~\ref{fig:AFM}e). The different signs of the resistance spikes for the different anti-parallel magnetization arrangments realized during switching suggests that these two phases are topologically distinct. Importantly, a recent related study based on conducting AFM CrSe--(Bi,Sb)$_{2}$Te$_{3}$ heterostructure showed that the proximity effect in these systems can crucially depend on the nature and symmetry of the termination-layer at the interface~\citep{yang2020termination}, which may be different for the top and bottom surfaces. 

In summary, new studies demonstrate that AFM materials can create an efficient interfacial exchange coupling when interfaced with MTI layers or TIs which can lead to enhancement of magnetic ordering. This shows that AFMs, just like FMs, can induce long-range coupling between two TIs through proximity effect in AFM--MTI heterostructures and high N\'{e}el ordering temperatures provided by the antiferromagnetic layers can induce higher temperature emergent phenomena such as QAHI or axion insulators. A remaining challenge is to realize strong proximity coupling in AFM--TI heterostructures with truly insulating AFMs with high N\'{e}el temperatures. 

\subsection{Emergent Phenomena}
\label{subsection:Emergent Phenomena}
Having reviewed the recent studies on magnetic proximity effect in both (ferromagnet) FM- and (antiferromagnet) AFM-based heterostructures, we now discuss the emergent phenomena enabled by proximity-induced magnetism in TIs, particularly the quantum anomalous Hall insulator (QAHI) and axion insulator states, discussed in Section~\ref{subsection:Quantum anomalous Hall insulator (QAHI) state and axion insulator state}. Apart from QAHI and axion insulators, ferromagnetic insulator--topological insulator--ferromagnetic insulator (FMI--TI--FMI) have also proven to be the playground of other exotic new phases, such as, topological Hall effect
(THE), which is also known as geometric Hall effect (GHE)~\citep{RN674,RN648,RN646}, planar Hall effect (PHE)~\citep{RN597}, and skyrmions~\citep{RN580,RN1083}, and these topics are also discussed below. The versatility of these heterostructures, also often makes it possible to achieve more than one type of such quantum phases in the same device~\cite{RN648, Mogieaao1669}. 

\subsubsection{Quantum anomalous Hall effect}
\label{subsubsection:Quantum Anomalous Hall Insulators}

\begin{figure}
\includegraphics[width=\columnwidth]{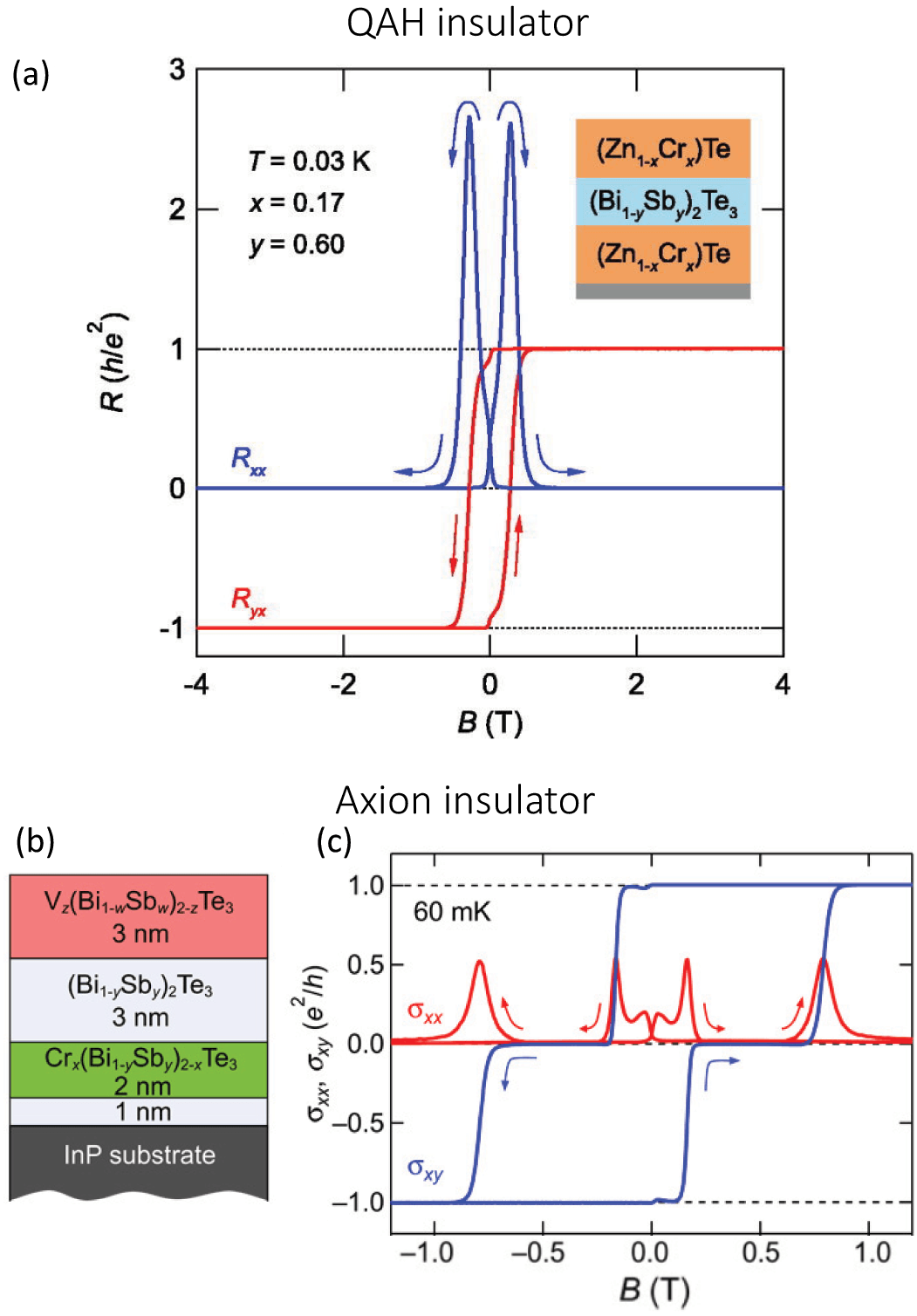}

\caption{Quantum anomalous Hall insulator and axion insulator. a) Magnetic
field $B$ dependence of $R_{xx}$ (blue) and $R_{yx}$ (red) at $T=0.03$~K for heterostructure of  (Zn$_{1-x}$Cr$_{x}$)Te--(Bi$_{1-y}$Sb$_{y}$)$_{2}$Te$_{3}$--(Zn$_{1-x}$Cr$_{x}$)Te shown schematically in the inset. Reprinted from Ref~\citep{RN569} with the permission of AIP publishing. b) Schematic structure of MBE-grown topological insulator (Bi$_{1-w}$Sb$_{w}$)$_{2}$Te$_{3}$ (BST) film modulation-doped with Cr and V used to demonstrate axion states in Figure (c). c) Magnetic field $B$ dependence of Hall conductivity (\textgreek{sv}$_{xy}$) and longitudinal conductivity (\textgreek{sv}$_{xx}$) of the Cr- and V-doped BST film at 60 mK. a) Reproduced from Appl. Phys. Lett. 115, 102403 (2019)~\citep{RN569}, with the permission of AIP Publishing \url{https://doi.org/10.1063/1.5111891}. b), c) Adapted from Science Advances, Vol. 3, no. 10, eaao1669~\citep{Mogieaao1669}. Distributed under a Creative Commons Attribution NonCommercial License 4.0 (CC BY-NC) \url{http://creativecommons.org/licenses/by-nc/4.0/}}{\label{fig:QAHandaxion}}
\end{figure}

 The QAHE in a MTI was initially predicted in 2010~\citep{yu2010quantized}, and Chang \emph{et al.}, experimentally confirmed QAHE in Cr-doped (Bi,Sb)$_{2}$(Te)$_{3}$ in 2013~\citep{chang2013experimental}. However, the QAHE has so far been realized in uniformly magnetic doped TIs only at sub-1~K temperatures~\citep{pan2020probing,mogi2015magnetic,RN694}, 1.4~K in stiochiometric MTIs~\citep{deng2020quantum} and up to a few kelvins with magnetic modulation doping of TIs~\citep{mogi2015magnetic}. As discussed in the Section~\ref{section: Introduction}, the limitation of operating temperature for QAHI to date appears to be due to disorder in magnetically doped systems, and difficulty of simultaneously optimizing magnetic exchange coupling and $T{_c}$ as well as topological inverted bandgap in intrinsic TIs.

The proximity-induced magnetism strategy~\citep{RN569,wei2013exchange} for acheiving the QAHE in a TI is the following: a TI is sandwiched between two FMI with parallel magnetizations which gap the TI surfaces though proximity-induced magnetism. The TI should be thin enough that the side surfaces are also gapped by confinement, but thick enough that the top and bottom surfaces have a hybridization gap smaller than the exchange gap. Theoretical studies have shown that when a TI is sandwiched between two FMI, it can undergo a transition into the quantum anomalous Hall phase
either from the topologically trivial phase or from the quantum spin Hall phase as a function of film thickness and strength of magnetic exchange~\citep{RN634}. Hou \emph{et al.} have discussed the realisation of QAHE at high temperature through DFT calculations for a van der Waals ferromagnet (vdWFM) monolayer-CrI$_3$ interfaced with Bi$_2$Se$_3$~\citep{Houeaaw1874}, and more detailed calculations predicted that when a Bi$_2$Se$_3$ film with thickness of 5 QL or more is sandwiched between two CrI$_3$ layers, it becomes a QAHI due to the competition between the exchange field from CrI$_3$ and interaction of the top and bottom surfaces of Bi$_2$Se$_3$ film~\citep{Houeaaw1874}. The high Curie temperature (up to 45~K) in CrI$_3$~\citep{huang2017layer} suggests the possibility of QAHI in CrI$_3$ - Bi$_2$Se$_3$ heterostructure at higher temperature than in uniformly doped TI materials. These predictions show the promise of magnetizing TSS using ferromagnetic materials through proximity effect in order to observe the QAHE as well as other phenomena at temperatures higher than a few kelvins~\citep{Houeaaw1874}.

The first attempt to achieve this experimentally was made by Mogi \emph{et al.}~\cite{mogi2015magnetic}, who demonstrated QAHI phase in ferromagnetic (Cr) modulation-doped (Bi$_{1-y}$Sb${_y}$)$_2$Te$_3$ (BST) thin films. In effect, this structure can be considered as an FMI--TI--FMI heterostructure, or as a modulated version of a dilute magnetically doped TI. Remarkably, they succeded in observing QAHE at temperatures up to 2~K, which is one of the highest  temperature QAHI-state obtained in FMI--TI heterostructures. Subsequently, there were several other reports of QAHE in similar heterostructures~\citep{RN1103, RN1138, RN1112, RN1122, RN648, Mogieaao1669}.

Watanabe \emph{et al.} first demonstrated QAHE in a FMI--TI--FMI heterostructure not involving modulation doping, by using an all-telluride structure~\citep{RN569}. Figure~\ref{fig:QAHandaxion}a), inset schematically shows the heterostructure (Zn$_{1-x}$Cr$_{x}$)Te--(Bi$_{1-y}$,Sb$_{y}$)$_2$Te$_3$--(Zn$_{1-x}$Cr$_{x}$)Te used by Watanabe \emph{et al.}~\citep{RN569}. An all-telluride heterostructure was deliberately chosen to engineer high-value of proximity-based exchange-coupling facilitated by overlap of Te-orbitals from both TI and FMI. Figure~\ref{fig:QAHandaxion}a) shows magnetotransport measurements on this structure at 30~mK; QAHI manifests as a perfect quantization of $R_{yx}=\frac{h}{e^2}$, while $R_{xx}$ approaches 0. This QAHI state persists in this heterostructure up to $\sim{0.1}~K$, comparable temperatures to dilute magnetically doped TIs. They conducted further investigations by varying material composition, which clearly showed the temperature range is limited by relatively low $T_c$ for the FMI, and the disorder in the heterostructure. As per our knowledge, this is the only demonstration of QAHE in a FMI--TI--FMI heterostructure not involving MTI. Obtaining QAHI at higher temperature requires further materials engineering and clear understanding of dissipative channels.

\subsubsection{Axion insulators and topological magnetoelectric effect}
\label{subsubsection: Axion Insulators}

\begin{figure*}
\includegraphics[width=\textwidth]{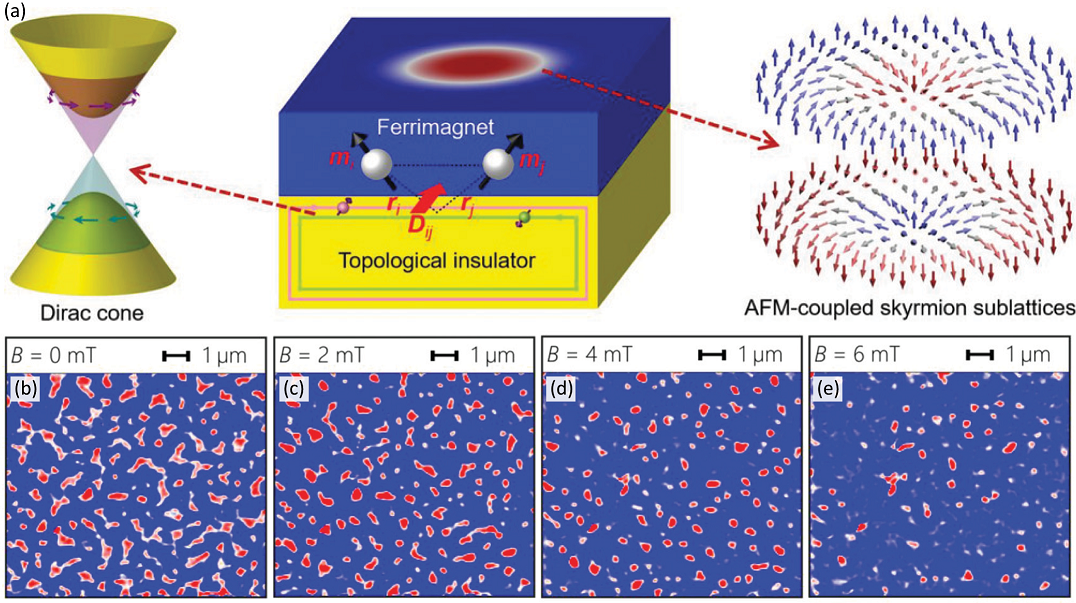}

\caption{Skyrmions in ferrimagnet (FiMI)--topological insulator (TI) heterostructures. a) Schematic of Dyzaloshinskii-Moriya interaction (DMI) and skyrmions in an FiMI--TI heterostructure. The middle panel shows a schematic of a FiMI-TI heterostructure; the strong
 spin-orbit coupling (SOC) in TIs combined with the structural inversion symmetry breaking by the interface produces the interfacial Dzyaloshinskii-Moriya interaction (DMI), which stabilizes the magnetic skyrmions in the ferrimagnet. The left panel show magnetic proximity-induced gap-opening at the Dirac point of topological surface states (TSS). The right panel shows antiferromagnetically coupled skyrmion sublattices in the ferrimagnet. b-e) Scanning transmission X-ray microscopy (STXM) images of skyrmions in GdFeCo--(Bi,Sb)$_2$Te$_3$ at room temperature. The colour scale indicates the magnetization, ranging from $-M{_z}$ (red) to $+M{_z}$ (blue) where $M_{z}$ is the saturation magnetization along the $z$-direction. A negative out-of-plane ($-z$) magnetic field ($B$) is used to saturate the magnetization to $-M_{z}$
first, and then the magnetic fields $B$ ranging from 0 to 8~mT in $+z$ are applied. STXM images are shown here at magnetic fields as indicated in each panel(b) 0~mT, (c) 2~mT, (d) 4~mT, and (e) 6~mT.Adapted from Adv. Mater. 2020, 32, 2003380~\citep{RN1083}. Distributed under Creative Commons license (CC BY 4.0)\url{https://creativecommons.org/licenses/by/4.0}}

{\label{fig:skyrmion1}}
\end{figure*} 

The axion insulator state has a bulk nontrivial topology but no longitudinal surface or edge conductivity~\citep{RN912,qi2008topological,RN912,morimoto2015topological}. 
As discussed in the Section~\ref{section: Introduction}, the axion insulator state can be realized in an appropriately magnetized 3D TI, and can be the lead candidate for observing  the topological magnetoelectric effect (TME).
As mentioned before, it is not possible  to  obtain  an  axion  insulator  state in  uniformly doped dilute magnetically doped topological insulators, as axion  insulator  requires  non-uniform magnetization of the crystal by definition~\citep{RN685,RN904}.
However, antiferromagnetism in MnBi$_{2}$Te$_{4}$ can enable a robust axion insulator phase with even layer number~\citep{liu2020robust}. 
Having said that, axion insulators more naturally appear in heterostuctures by creating antiparallel magnetizations via proximity effect~\citep{Mogieaao1669}, as shown schematically in Figure~\ref{fig:Emergent-phenomena}c). 
Theoretical studies have also predicted an axion insulator state in a FM--TI--AFM heterostructure, that can be observed in a wide range of external magnetic field~\citep{RN582}. 
Moreover, Hou \emph{et al.} has recently shown CrI$_3$--Bi$_2$Se$_3$--MnBi$_2$Se$_4$ heterostructures yield a robust axion insulator phase through DFT calculation and four-band model studies. Their finding suggested that observation of an axion insulator in this heterostructure is a result of opposite-sign exchange fields into Bi$_2$Se$_3$ surface states induced by CrI$_3$ and MnBi$_{2}$Se$_{4}$ over layers~\citep{RN622}. 

Mogi \emph{et al.}~\citep{Mogieaao1669} have demonstrated a robust axion insulator state in a wide range of magnetic and electric fields by tailoring a tricolor structure of a magnetic thin film. As shown in Figure~\ref{fig:QAHandaxion}b); a non-magnetic TI ((Bi,Sb)$_2$Te$_3$) was interfaced with Cr-doped and V-doped (Bi,Sb)$_2$Te$_3$ with different coercive magnetic fields ($B_c$) and the non-magnetic layer ensures the magnetic decoupling between the two magnetic layers. Figure~\ref{fig:QAHandaxion}c) shows the transport measurements of Cr-V doped BST at 60~mK. The different coercive fields result in a double-step magnetization reversal, reflected in the double step hysteresis loop of $\sigma_{xy}(B)$, with anti-parallel magnetization of the layers realized between two coercive fields. The QAHI is observed at high $B$, while the hallmark of the axion insulator, zero Hall conductivity plateaus (ZHP) with simultaneously zero longitudinal conductivity, was observed  in the anti-parallel state~\citep{Mogieaao1669}: \textgreek{sv}$_{xx}$ is almost zero between \textgreek{sv}$_{xy}$ transitions where the ZHP width is observed in a range of 0.19~T\ensuremath{\le}B\ensuremath{\le}0.72~T. For comparison, V-V doped and Cr-Cr doped BST heterostructurs were also measured in this study and surprisingly, the ZHP are seen in the transport measurements of these heterostructures as well, possibly as a result of unintentional differences in $B_c$~\citep{Mogieaao1669} of the top and bottom magnetic layers. In a similar Cr-doped (Bi,Sb)$_2$Te$_3$--(Bi,Sb)$_2$Te$_3$--Cr-doped (Bi,Sb)$_2$Te$_3$ heterostructure Allen \emph{et al.} used microwave impedance microscopy (MIM) to directly image the conducting edge modes of the QAHI and absence of conducting edge modes in the intervening axion insulator state occuring at the
boundary between two QAHI states of opposite chirality~\citep{Allen14511}.

\begin{figure*}[th]
\includegraphics[width=\textwidth]{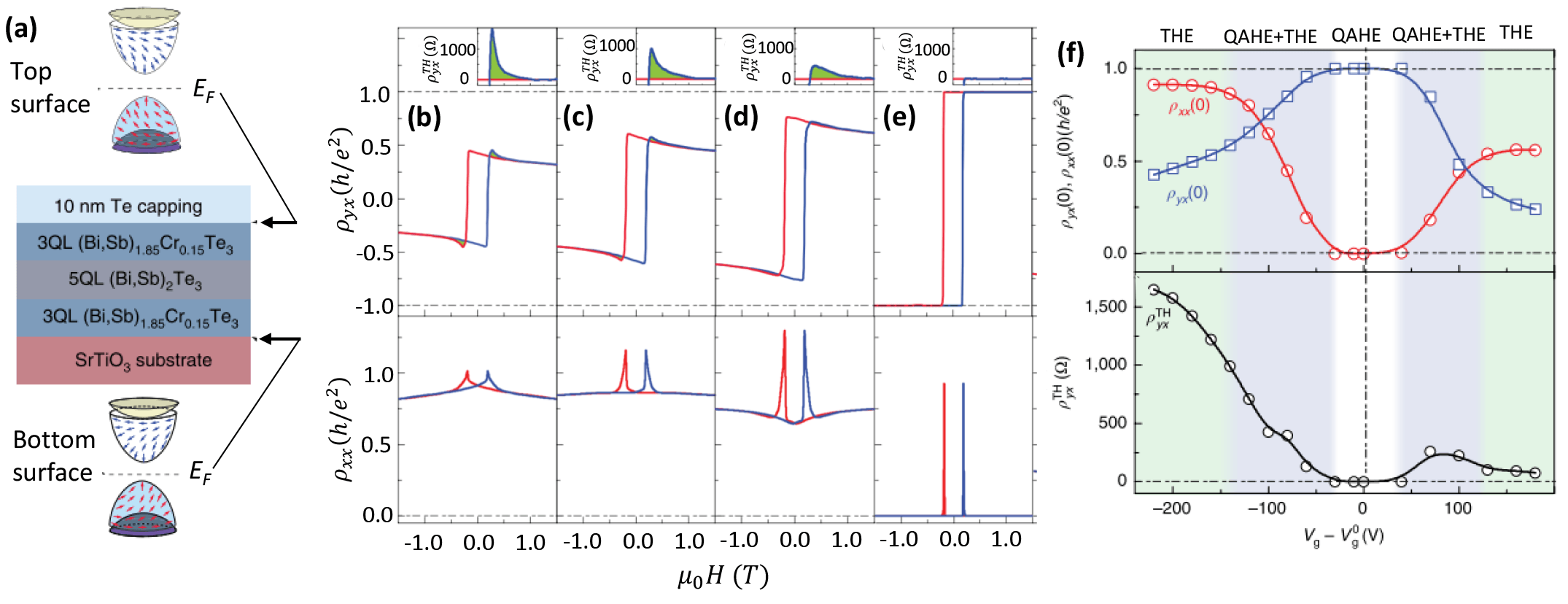}

\caption{Coexistent topological Hall effect (THE) and quantum anomalous Hall effect (QAHE). a) Schematic of the magnetic topological insulator (MTI)--topological insulator (TI)--magnetic topological insulator (MTI) sandwich heterostructure. When $T < T_{c}$, an exchange gap opens at the Dirac points of the top and bottom surfaces. Blue (red) arrows represent the spin orientations of the top (bottom) section of gapped Dirac surface states (SSs). b-e) Gate-induced THE in MTI--TI--MTI sandwich heterostructures. Magnetic-field ($\mu_{0}H$) dependence of the Hall resistance $\rho_{yx}$ (top) and the longitudinal resistance $\rho_{xx}$ (bottom) at different gate voltages relative to the charge neutrality point $V_{g}-V_{g}^{0}$ of -220 V (b), -140 V (c), -100 V (d), and 0 V (e) at $T= 30$~mK. The sample shows perfect QAHE when $V_{g}-V_{g}^{0}=0$~V. When $V_{g}$ is tuned away from $V_{g}^{0}$, $\rho_{yx}$ deviates from the quantized value ($h/e^{2}$) and shows hump (green shaded feature). f) Upper panel shows gate dependence of the Hall resistance $\rho_{yx}$(0) (empty blue squares) and the longitudinal resistance $\rho_{xx}$(0) (empty red circles) at zero magnetic field and $T=30$~mK. Lower panel shows gate dependence of the topological Hall (TH) resistance $\rho_{yx}^{TH}$ at T = 30 mK, the regions of concurrence of the quantum anomalous Hall (QAH) and TH effects are shaded light blue. Reprinted by permission from Rightslink Printable License: Springer Nature Nature Materials Concurrence of quantum anomalous Hall and topological Hall effects in magnetic topological insulator sandwich heterostructures~\citep{RN648}, Jue Jiang, Di Xiao, Fei Wang, Jae-Ho Shin, Domenico Andreoli, Jianxiao Zhang, Run Xiao, Yi-Fan Zhao, Morteza Kayyalha, Ling Zhang, Ke Wang, Jiadong Zang, Chaoxing Liu, Nitin Samarth, Moses H. W. Chan and Cui-Zu Chang,  \copyright 2020, The Author(s), under exclusive licence to Springer Nature Limited].}{\label{fig:skyrmion2}}
\end{figure*} 

\subsubsection{Skyrmions}
\label{subsubsection: Skyrmions}

Magnetic skyrmions are chiral spin structures exhibiting non-trivial topology in real space~\citep{roessler2006spontaneous,jonietz2010spin,romming2013writing,mochizuki2014thermally, RN580}, as illustrated in Figure~\ref{fig:skyrmion1}a). They can be energetically stabilized by the Dzyaloshinskii-Moriya interaction (DMI) which can occur naturally at the interface between a magnet and a strong spin-orbit coupled material, making magnetic material--topological insulator (MM--TI) heterostructures natural candidates for observing stable skyrmions. Energetically stable skyrmions could be useful as non-charge-based carriers of information in spintronic devices and memories, owing to their topological protection, adding potential additional functionalities to MM--TI heterostructures.

 Wu \emph{et al.} directly imaged skyrmions for the first time in a ferrimagnet-TI heterostructure of GdFeCo--(Bi,Sb)$_2$Te$_3$ at room temperature~\citep{RN1083}. The strong interfacial DMI, resulting from spin-orbit coupling (SOC) of TI and breaking of inversion-symmetry at the interface, facilitates superexchange interaction between the magnetic moments of GdFeCo, stabilizing magnetic skyrmions (as shown in Figure~\ref{fig:skyrmion1}a). Figure~\ref{fig:skyrmion1}b-e) shows a colour scale plot of the magnetization of the same structure, obtained through scanning transmission X-ray microscopy (STXM). The magnetization was first saturated in the -z direction with a negative out-of-plane magnetic field, and Figure~\ref{fig:skyrmion1}b-e) show STXM data for progressively more positive applied magnetic fields. With increasing magnetic field, the strip domains with $-M_{z}$ continue to shrink, forming the skyrmion states above $B=4$~mT. Skyrmionium, a soliton made of two skyrmions with opposite winding numbers, has also been imaged in FM-MTI heterostructures of Ni$_{80}$Fe$_{20}$--Cr-doped (Bi,Sb)$_2$Te$_3$ using X-ray photoemission electron microscopy (XPEEM)~\citep{RN580}.

Skyrmions exhibit a signature in electronic transport of non-zero spin chirality known as the topological Hall effect (THE)~\citep{RN648}. Similar to the AHE which arises as a consequence of non-zero Berry curvature in momentum space, the topological Hall effect is a consequence of Berry curvature in real space~\citep{RN648}. The experimental signature of THE is an additional current ``hump" in the magnetic field-dependent anomalous Hall conductivity generally occurring near the magnetization reversal~\citep{nagaosa2013topological}.
Previously, the THE has been observed in metallic systems~\citep{neubauer2009topological,kanazawa2011large,matsuno2016interface,ohuchi2018electric} and in magnetically doped TI films and heterostructures~\citep{yasuda2016geometric,liu2017dimensional}. Qing Lin He \emph{et al.} have successfully tuned the sign of THE by changing the field-cooling direction~\citep{RN646}.

Generally the QAHE is observed in the bulk insulating regime~\citep{chang2013experimental,V_doped_QAHE_chang2015high,mogi2015magnetic} while the THE requires carriers in the bulk~\citep{yasuda2016geometric,liu2017dimensional}. Interestingly, in a recent study, researchers have observed QAHE and THE in a single sample by tuning the chemical potential of a  FMI--TI heterostructure. Figure~\ref{fig:skyrmion2}a) shows the heterostructure under study: a non-magnetic TI layer (5 QL (Bi,Sb)$_{2}$(Te,Se)$_{3}$) is sandwiched between two MTIs (3 QL (Bi,Sb)$_{1.85}$(Cr)$_{0.15}$(Te,Se)$_{3}$) in a field-effect geometry for carefully tuning the chemical potential~\citep{RN648}. In order to observe both effects (QAHE +THE) in one sample three conditions should be met: (a) breaking the TRS as a prerequisite for both QAHE and THE, (b) strong Dzyalshinskii-Moriya interaction (DMI) as an essential requirement for the THE, and (c) Tuning the chemical potential of the top and the bottom surfaces simultaneously near the magnetization exchange gaps, which is required for the QAHE~\citep{RN648}. 

Figure~\ref{fig:skyrmion2}b-e show the result. In Figure~\ref{fig:skyrmion2}e), at the charge neutrality point, \emph{i.e.} $(V_{g}-V_{g}^0)=0$~V the sample shows a perfect QAHE as the Hall curves overlap during the magnetic field sweeps when magnetic field is greater than the coercive field. When the chemical potential is tuned such that the QAHE and THE coexist (Figure~\ref{fig:skyrmion2}b-d), an extra Hall current (green shaded area) is induced by the emergent gauge field as spin polarized carriers adiabatically pass through real space topological spin textures. Figure~\ref{fig:skyrmion2}f) summarizes the gate voltage-dependence of the zero-field longitudinal and Hall resistivities $\rho_{xx}$(0) and $\rho_{yx}$(0) (top panel) as well as the additional Hall resistivity due to the THE $\rho_{yx}^{TH}$ (bottom panel) at $T$ = 30mK. The sample is in QAH state in the white shaded area where $-30V\leq V_{g}-V_{g}^{0}\leq+40V$. Outside this region of gate voltage the current flows through both the dissipationless chiral edge channels through the dissipative bulk channel. Therefore, when $-140V\leq V_{g}-V_{g}^{0}\leq-30V$ and $+30V\leq V_{g}-V_{g}^{0}\leq+120V$ the sample shows imperfect QAHE (the suppression of $\rho_{xx}(0)$ indicates some current flows in the edge state) as well as a finite THE coexisting in this region~\citep{RN648}. The THE is an attractive technique to detect skyrmions in MM--TI heterostructures. However, it has two drawbacks: (i) it is an indirect technique, (ii) it often appears along with other magnetotransport phenomena like QAHE and has a relatively weaker signature; therefore direct imaging of skyrmions through XPEEM or STXM are comparatively more reliable.

\subsubsection{Magnon-mediated superconductivity}
\label{subsubsection:Magnon-mediated superconductivity}

Heterostructures of TIs and magnetic materials (MMs) are also predicted to present the ideal framework to realize another new, and counterintuitive quantum phase, magnon-mediated superconductivity, where superconductivity is assisted by magnetism. Magnons are quasiparticles resulting from collective excitations of magnetic
moments in MMs. As magnons are also bosons, they can
mediate attractive interaction between electrons which can result
in superconductivity. Heterostructures of TIs and magnetic insulators (MIs) are predicted to be ideal for observing such a phenomenon.
Kargarian \emph{et al.} predicted that in a ferromagnetic insulator--topological insulator (FMI--TI) heterostructure,
ferromagnetic fluctuations can lead to an Amperean type attractive interaction
between electrons with parallel momenta at the TSS and can lead to
an instability towards a superconducting state with finite-momentum
Cooper pairs~\citep{kargarian2016amperean}. Hugdal \emph{et al.}~\citep{RN629}
and Erlandsen \emph{et al.}~\citep{RN620} showed that both BCS and
Amperean type attractive interaction can happen in both FMI--TI and antiferromagnetic insulator--topological insulator (AFMI--TI) heterostructures depending on various
parameters. However, according to Erlandsen \emph{et al.} an uncompensated antiferromagnetic interface is better than a ferromagnetic interface in inducing magnon-mediated superconductivity, as the interaction strength in the former can be enhanced by asymmetric coupling of the TI with the two sublattices of the antiferromagnet~\citep{RN620}. 

A spontaneous time reversal symmetry-broken superconducting state has already been observed in a heterostructure of FM Ni and semimetal Bi~\citep{xin2015possible}, which is now known to have topological properties~\citep{schindler2018higher}. Kargarian  \emph{et al.} predicted, a magnon-mediated superconductivity should be observable in a heterostructure with Bi$_{2}$Se$_{3}$ or Bi$_{2}$Te$_{3}$ as TI and Ni or EuS as the ferromagnet with a superconducting transition temperature of 1 K~\citep{kargarian2016amperean}. However, no experiments in such systems have detected superconductivity yet, and this area appears ripe for future developments as the the science of MM--TI heterostructrures evolves.

\section{Conclusion and Outlook}
\label{section:Conclusion and Outlook}
In this review, we have focused on the emerging field of magnetic materials interfaced with topological insulators in engineered heterostructures with the goal of establishing robust high temperature quantum phases such as the QAHI and axion insulator. The last few years has seen rapid advancements in understanding the mechanisms of the proximity effect, optimizing fabrication and measurement techniques, discovering new ferromagnetic and topological materials, and achieving and fine-tuning various emergent phenomena. The next challenge is to develop practical applications, which would require understanding, designing, and developing novel devices based on MM--TI heterostructures.

\subsection{Future applications}
\label{subsection:Future applications}
One of the grand challenges of our time is to find an alternative for the Si-based complementary metal oxide semiconductor (CMOS) technology which has almost reached its limits of in terms of miniaturization, speed, and cost-effectiveness~\citep{moore2020international, khan2018science}. The end of CMOS miniaturization will be accompanied by a drastic slowdown in the increases in energy efficiency in information processing, which will constrain the continued growth in worldwide computation capacity~\citep{andrae2015global}. 

Among the various proposed routes to faster, more energy-efficient electronics, spintronics is one of the most promising. Spintronics exploits the electron's spin for both data storage and processing of logic operations~\citep{wang2018spintronic, ningrum2020recent, joshi2016spintronics}. Logical states (0 and 1) are represented by magnetization configurations, e.g. polarization of the electron spins in a conductor, the direction of magnetization of a ferromagnet, or topological charge of a skyrmion. These states are generally isoenergetic, which in principle can allow switching between states with very little wasted energy. The state can be stabilized by an energy barrier to become very long lived (as is well known from non-volatile magnetic storage) and thus can naturally combine logic and storage functions in a single technology. 

A fully functional spintronic technology requires an efficient framework for spin-generation, spin-transport, spin-manipulation, and spin-detection~\citep{lin2019two}. It is even more preferable if all these functionalities can be achieved in the same materials system to enable highly-dense devices. The unique emergent phenomena in magnetic topoological insulators have the capability to achieve the versatile functionalities demanded by spintronics.

One of the issues with first-generation spintronics devices was reliance on external magnetic field for spin-operations, which is neither energy- or time-efficient, nor practical. It is preferable to utilize materials with high spin-orbit coupling, and manipulate spins with a charge current. Due to the high spin-orbit coupling in a TI, and spin-polarised nature of the TSS, the bilayer FM-TI heterostructures has already shown excellent results in terms of current-induced spin-injection and magnetization switching~\citep{anthony2018molecular, fert2019spintronics, RN878, han2018quantum}, even with room temperature functionality~\citep{han2017room, RN696}. Nonetheless, a major issue with charge-current-induced spin-manipulation is the Joule heating associated with the current. Here the dissipation-free edge states of QAHI offer a significant advantage: the edge states are 100$\%$ spin-polarised, and carry charge currents without dissipation. QAHI has been realized in devices with mm-size length-scale~\citep{kou2015metal}, and there is no theoretical upper limit on the mean free path of such edge states, making these channels ideal for charge-spin conversion and spin-transport. Moreover, spin-to-charge readout in these devices can be achieved  by reorienting magnetic moments of FMI, which reverses the chirality (charge current direction) of the edge states. Proposed memory devices based on QAHI systems also offer added advantages of constant readout voltages even in the presence of imperfections and can help making integrated circuits simpler by alleviating the usage of voltage comparators~\citep{fujita2011topological}. In fact, manipulation of chiral edge states by rewriting of magnetic domains has already been demonstrated in QAHI devices~\citep{yasuda2017quantized}. 

A major challenge in the development of topological spintronic devices will be to utilize the spin-polarized current carried by the QAHI to alter the magnetization of another QAHI. Since no current flows in the bulk of the QAHI, it is not clear how the spin-polarized edge current will affect the magnetization of the bulk insulator. Again, interfaces are likely to be important; the edge current of a QAHI could be used to manipulate a FM metal proximity-coupled to the bulk of another QAHI.  

Key to a successful spintronics technology will be to reduce the energy required to switch the spin configuration. In this respect, proximity-induced QAHI states are attractive, as the proximity coupling is an electronic phenomenon and can be manipulated by e.g. external gates~\citep{RN688,RN648}. Another route to manipulating the spin configuration has been to use multiferroic materials, which possess coexisting and coupled ferroelectric and ferromagnetic properties~\citep{spaldin2019advances}, enabling a magnetic polarization through application of electric field. Multiferroics can allow the ferromagnetic transition temperature or anisotropy to be manipulated electrically, lowering the energy barrier to switch from one spin configuration to another~\citep{trassin2015low}. Axion insulators naturally exhibit the topological magnetoelectric effect, a material-independent, quantized high value of magnetoelectric coupling constant, and hence are strong candidates for spintronic applications. Furthermore, QAHI and axion insulators can be realized in the same materials system: a trilayer heterostructure of magnetic material-topological insulator-magnetic material (MM--TI--MM) can be continuously tuned from QAHI to axion insulator~\citep{Mogieaao1669}, and a large area trilayer device can be split into multiple devices with different domains hosting QAHI states with different chiralities as well as axion insulator states. Hence, it is possible to create a densely packed integrated device in such heterostructures. A major challenge in the near future will be to demonstrate the topological magnetoelectric effect, and use it to electrically manipulate a spin configuration. 

Lastly, skyrmions have garnered recent interest as a spin configuration for memory and logic due to their compactness (<10 nm diameter), quantized topological charge, and high-velocity (100 m/s)~\citep{fert2019spintronics, hsu2017electric, fert2013skyrmions}. Hence skyrmions obtained in TI-FM heterostructure have a high potential to be integrated with QAHI or axion-insulator based devices as information or memory elements.

There are several other applications of such heterostructures beyond the area of spintronics. Already QAHI has been demonstrated with a quantization within 2 ppm of the ideal value of $\frac{h}{e^2}$. Upon further improvement (ppb) and room temperature realization, QAHI could provide a primary standard of resistance in metrology instead of the quantum Hall effect, which requires low temperature and high magnetic field~\citep{poirier2009resistance, rigosi2019quantum}. Even the anomalous Hall effect in MM--TI heterostructures has huge potential in highly sensitive and low-power Hall sensors utilized in the fields from medical science to marine technology~\citep{ning2020ultra, wang2020anomalous}.


\makeatletter
\renewcommand*{\thefootnote}{\fnsymbol{footnote}}
\newcommand\footnoteref[1]{\protected@xdef\@thefnmark{#1}\@footnotemark}
\makeatother
\begingroup
\squeezetable
\begin{table*}[!]\centering
\caption{Recent measurements on topological insulator-magnetic material heterostructures (mainly, 2017-present)\label{tab:Details-of-measurements}}
\begin{tabular}{|l|l|l|l|l|l|l|} \hline
\textbf{Ref\#} & \begin{tabular}{l}
    \textbf{(A)FMI / }\\\textbf{FMM / }\\\textbf{FiMI\footnote{Ferrimagnetic insulator} / }\\\textbf{MTI}
\end{tabular} & \textbf{FM-mat} & \textbf{TI-mat} & \begin{tabular}{l}
     \textbf{QAHE}\\\textbf{Temp}
\end{tabular} & \begin{tabular}{l}
    \textbf{Mag}\\\textbf{Temp}
\end{tabular} & \begin{tabular}{l}
    \textbf{Measurement}\\\textbf{Technique}
\end{tabular} \\\hline
\multicolumn{7}{|c|}{\textit{\textbf{Magnetic Insulator - TI Heterostructures}}} \\\hline
~\citep{RN941} & FMI & CoFe$_{2}$O$_{4}$ & (Bi$_{1-y}$Sb$_{y}$)$_{2}$Se$_{3}$ & - & 2 K \footnote{\label{note:*}Highest measured temperature} & Magneto-transport \\\hline
~\citep{RN663} & FMI & CrGeTe$_{3}$ & (Bi$_{1-y}$Sb$_{y}$)$_{2}$Te$_{3}$ & - & 150 K & Magnetometry \\\hline
~\citep{RN581} & FMI \footnote{van der Waals heterostructure from exfoliated crystals} & CrGeTe$_{3}$ & BiSbTeSe$_{2}$ & 1.5 K & 60 - 80 K & Magneto-transport \\\hline
~\citep{RN583} & FMI & CrGeTe$_{3}$ & (Bi$_{1-y}$Sb$_{y}$)$_{2}$Te$_{3}$ & - & 65 K & Magneto-transport \\\hline
~\citep{RN637} & FMI & CrGeTe$_{3}$ & (Bi$_{1-y}$Sb$_{y}$)$_{2}$Te$_{3}$ & - & 80 K & \begin{tabular}{l}
      Neutron reflectometry, \\magneto-transport,\\ magnetometry 
 \end{tabular} \\\hline
~\citep{pan2020observation} & AFMI & Cr$_{2}$O$_{3}$ & Cr$_{x}$(Bi$_{1-y}$Sb$_{y}$)$_{2-x}$Te$_3$ & 0.02 K & 297.3 K & 	Neutron Reflectometry\\\hline 
~\citep{RN598} & FMI & EuS & Bi$_{2}$Se$_{3}$ & - & 300 K \footnoteref{\ref{note:*}} & \begin{tabular}{l}
      Magnetometry,\\neutron reflectometry
 \end{tabular} \\\hline
~\citep{RN688} & FMI & EuS & (Bi$_{1-y}$Sb$_{y}$)$_{2}$Te$_{3}$ & - & - & \begin{tabular}{l}
      Neutron reflectometry,\\magneto-transport
 \end{tabular} \\\hline
~\citep{RN630} & FMI & EuS & Bi$_{2}$Se$_{3}$ & - & $\sim$ 60 K & Magnetometry \\\hline
~\citep{RN692} & FMI & EuS & \begin{tabular}[c]{@{}l@{}}(Bi$_{1-y}$Sb$_{y}$)$_{2}$Te$_{3}$\\ Sb2Te3\end{tabular} & - &$\sim$ 60 K & \begin{tabular}{l}
    Magneto-transport,\\magnetometry
 \end{tabular} \\\hline
~\citep{RN597} & FMI & EuS & (Bi$_{1-y}$Sb$_{y}$)$_{2}$Te$_{3}$ & - & - & Magneto-transport \\\hline
~\citep{RN693} & FMI & EuS & \begin{tabular}[c]{@{}l@{}}Bi$_{2}$Se$_{3}$\footnote{\label{note:primary}Primary material}\\ V$_{x}$(Bi$_{1-y}$Sb$_{y}$)$_{2-x}$Te3\\ Bi\end{tabular} & - & 16 K & \begin{tabular}{l}
      Muon spin spectroscopy,\\ARPES
 \end{tabular} \\\hline
 ~\citep{RN1125} & FMI & EuS & Bi$_{2}$Se$_{3}$ & - & 19 K & \begin{tabular}{l}
    Magneto-transport,\\Magnetometry
 \end{tabular} \\ \hline
~\citep{RN586} & FMI & Ga$_{1-x}$Mn$_{x}$As & Bi$_{2}$Se$_{3}$ & - & 147 K & \begin{tabular}{l}
    Magneto-transport,\\magnetometry
 \end{tabular} \\\hline
~\citep{RN647} & FMI & Ga$_{1-x}$Mn$_{x}$As & (Bi$_{1-y}$Sb$_{y}$)$_{2}$(Se$_{1-z}$Te$_{z}$)$_{3}$ & - &$\sim$ 50 K & Magneto-transport \\\hline
~\citep{RN592} & FMI & LaCoO$_{3}$ & Bi$_{2}$Se$_{3}$ & - &$\sim$ 70 K & Magneto-transport \\\hline
~\citep{RN589} & AFMI & NiO & Bi$_{2}$Se$_{3}$ & - & - & Magneto-transport \\\hline
~\citep{Tange1700307} & FiMI & Tm$_{3}$Fe$_{5}$O$_{12}$ & (Bi$_{1-y}$Sb$_{y}$)$_{2}$Te$_{3}$ & - & 400 K \footnoteref{\ref{note:*}} & \begin{tabular}{l}
    Magneto-transport,\\PCAR spectroscopy
 \end{tabular} \\\hline
~\citep{RN664} & FiMI & Tm$_{3}$Fe$_{5}$O$_{12}$ & Bi$_{2}$Se$_{3}$ & - &$\sim$ 200 K & \begin{tabular}{l}
      Magneto-transport,\\magnetometry
 \end{tabular} \\\hline
~\citep{RN1004} & FMI & Y$_{3}$Fe$_{5}$O$_{12}$ & (Bi$_{1-y}$Sb$_{y}$)$_{2}$Te$_{3}$ & - & - & \begin{tabular}{l}
      Magnetometry,\\magneto-transport
 \end{tabular} \\\hline
~\citep{RN645} & FMI & Y$_{3}$Fe$_{5}$O$_{12}$ & Bi$_{2}$Se$_{3}$ & - & 180 K & Magnetometry \\\hline
~\citep{RN675} & FMI & Y$_{3}$Fe$_{5}$O$_{12}$ & Bi$_{2}$Se$_{3}$ & - & 30 K & Magneto-transport \\\hline
~\citep{RN1109} & FiMI & Y$_{3}$Fe$_{5}$O$_{12}$ & Bi$_{2}$Se$_{3}$ & - & - & Magnetometry
\\\hline
~\citep{RN939} & \begin{tabular}{l}
      FMI\\FiMI
 \end{tabular} & \begin{tabular}[c]{@{}l@{}}Y$_{3}$Fe$_{5}$O$_{12}$\\ Tm$_{3}$Fe$_{5}$O$_{12}$\end{tabular} & Bi$_{2}$Se$_{3}$ & - & 180 K & Magneto-transport \\\hline
~\citep{RN569} & FMI & Zn$_{1-x}$Cr$_{x}$Te & (Bi$_{1-y}$Sb$_{y}$)$_{2}$Te$_{3}$ & 0.1 K &$\sim$ 40 K & Magneto-transport \\\hline
\multicolumn{7}{|c|}{\textit{\textbf{Magnetic TI - TI Heterostructures}}} \\\hline
~\citep{RN639} & FTI & Cr$_{x}$(Sb$_{2-x}$Te$_{3}$) & Dy$_{x}$(Bi$_{2-x}$Te$_{3}$) & - & \begin{tabular}{l}
      17 K (Dy),\\70 K (Cr)
 \end{tabular} & Magnetometry \\\hline
~\citep{yasuda2016geometric} & FTI & Cr$_{x}$(Bi$_{1-y}$Sb$_{y}$)$_{2-x}$Te$_{3}$ & (Bi$_{1-y}$Sb$_{y}$)$_{2}$Te$_{3}$ & - &$\sim$ 18 K & Magneto-transport \\\hline
~\citep{RN685} & FTI & Cr$_{x}$(Bi$_{1-y}$Sb$_{y}$)$_{2-x}$Te$_{3}$ & (Bi$_{1-y}$Sb$_{y}$)$_{2}$Te$_{3}$ & 3 K & 40 K & Magneto-transport \\\hline
~\citep{yasuda2017quantized} & FTI & Cr$_{x}$(Bi$_{1-y}$Sb$_{y}$)$_{2-x}$Te$_{3}$ & (Bi$_{1-y}$Sb$_{y}$)$_{2}$Te$_{3}$ & 0.5 K & - & \begin{tabular}{l}
    Magneto-transport,\\magnetometry
 \end{tabular} \\\hline
~\citep{RN648} & FTI & Cr$_{x}$(Bi$_{1-y}$Sb$_{y}$)$_{2-x}$Te$_{3}$ & (Bi$_{1-y}$Sb$_{y}$)$_{2}$Te$_{3}$ & 0.6 K & 19 K & Magneto-transport \\\hline
~\citep{Mogieaao1669} & FTI & \begin{tabular}[c]{@{}l@{}}Cr$_{x}$(Bi$_{1-y}$Sb$_{y}$)$_{2-x}$Te$_{3}$\\ V$_{z}$(Bi$_{1-w}$Sb$_{w}$)$_{2-w}$Te$_{3}$\end{tabular} & (Bi$_{1-y}$Sb$_{y}$)$_{2}$Te$_{3}$ & 0.3 K & 20 K \footnoteref{\ref{note:*}} & Magneto-transport \\\hline
\multicolumn{7}{|c|}{\textit{\textbf{Magnetic Metal - TI Heterostructure}}} \\\hline
~\citep{RN611} & AFMM & CrSb & Cr$_{x}$(Bi$_{1-y}$Sb$_{y}$)$_{2-x}$Te$_{3}$ & - & 35 K & \begin{tabular}{l}
      Neutron reflectometry, \\magnetometry, \\magneto-transport 
 \end{tabular} \\\hline
~\citep{RN694} & AFMM & CrSb & (Bi$_{1-y}$Sb$_{y}$)$_{2}$Te$_{3}$ & - & 90 K & Magneto-transport \\\hline
~\citep{yang2020termination} & AFMM & CrSe & (Bi$_{1-y}$Sb$_{y}$)$_{2}$Te$_3$ & - & 120 K & \begin{tabular}{l}
     Magneto-transport,\\ Magnetometry\\, Neutron reflectometry
\end{tabular}\\\hline
~\citep{RN699} & FMM & Cr$_{2}$Te$_{3}$ & Cr$_{x}$Sb$_{2-x}$Te$_{3}$ & - & 15 K & Magnetometry \\\hline
~\citep{RN674} & FMM & Cr$_{2}$Te$_{3}$ & Bi$_{2}$Te$_{3}$ & - & 25 K & Magneto-transport \\\hline
~\citep{RN683} & FMM & Co$_{40}$Fe$_{40}$B$_{20}$ & Bi$_{2}$Se$_{3}$ & - & - & - \\\hline
~\citep{RN590} & FMM & \begin{tabular}{l}
      Co$_{40}$Fe$_{40}$B$_{20}$,\\Co$_{55}$Fe$_{45}$
 \end{tabular} & \begin{tabular}{l}
      Bi$_{2}$Te$_{3}$,\\ Bi$_{2}$Se$_{3}$
 \end{tabular} & - & - & Magnetometry \\\hline
~\citep{RN599} & FMM & Fe & Sb$_{2}$Te$_{3}$ & - & - & - \\\hline
~\citep{RN1078} & FMM & Fe$_{3}$GeTe$_{2}$ & Bi$_{2}$Te$_{3}$ & - & 400 K & \begin{tabular}{l}
    Magneto-transport, \\Magnetometry
\end{tabular}\\\hline
~\citep{RN591} & FMM & Fe$_{3}$Si, Co$_{2}$FeSi & Bi$_{2}$Te$_{3}$ & - & 10 K \footnoteref{\ref{note:*}} & Magnetometry \\\hline
~\citep{RN642} & FMM & FeSe$_{x}$ & Bi$_{2}$Se$_{3}$ & - & - & Magnetometry \\\hline
~\citep{RN646} & AFMM & MnTe & (Bi$_{1-y}$Sb$_{y}$)$_{2}$Te$_{3}$ & - & $\sim$ 240 K & Magneto-transport \\\hline
~\citep{RN816} & FMM & MnTe & \begin{tabular}{l}
      Bi$_{2}$Te$_{3}$\footnoteref{\ref{note:primary}}\\(Bi$_{1-y}$Sb$_{y}$)$_{2}$Te$_{3}$
 \end{tabular} & - & $\sim$ 130 K & Magneto-transport \\\hline
~\citep{RN659} & FMM & Ni & Bi$_{2}$Se$_{3}$ & - & 4 K \footnoteref{\ref{note:*}} & Magnetometry \\\hline
~\citep{RN567} & FMM & Ni & Bi$_{2}$Se$_{3}$ & - & 300 K \footnoteref{\ref{note:*}} & Magneto-transport \\\hline
~\citep{RN593} & FMM & Ni$_{80}$Fe$_{20}$ & Bi$_{2}$Se$_{3}$ & - & 300 K \footnoteref{\ref{note:*}} & Magnetometry \\\hline
~\citep{RN580} & FMM & Ni$_{81}$Fe$_{19}$ & (Cr$_{x}$Sb$_{1-x}$)$_{2}$Te$_{3}$ & - & 100 K & Electron microscopy \\\hline
~\citep{RN627} & FMM & SrRuO$_{3}$ & (Bi$_{1-y}$Sb$_{y}$)$_{2}$Te$_{3}$ & - & 150 K & Magneto-transport\\
 \hline
\end{tabular}
\end{table*}
\endgroup

\renewcommand*{\thefootnote}{\arabic{footnote}}

\subsection{Materials challenges}
In order to achieve the full potential of the applications mentioned in the previous section, it is important to realize emergent phenomena like QAHE and topological magnetoelectric effect at room temperature, which remains a significant challenge. There appears to be no fundamental physical limitation to the temperatures for these effects, hence it is a matter of finding the ideal materials combinations for MM--TI--MM heterostructures from the continuously expanding library of FMI and TI materials, and creating heterostructures with minimal disorder. 

Traditionally, such heterostructures have been synthesized by MBE, which imposes a constraint on the number of such combinations, as MBE growth is extremely sensitive to lattice parameters of the two materials and growth conditions. Therefore, other synthesis techniques which are more versatile need to be explored.

In this context, proximity-coupled MM--TI structures are poised to take advantage of another recent materials revolution: the advent of van der Waals heterostructures~\citep{geim2013van}. This technique relies only on van der Waals attraction of adjacent layers, removing the constraint of lattice matching, and hence can enable new materials combinations in the search for room temperature QAHI and axion insulator~\citep{RN581}. Intelligent modification of this technique can even lead to integration of non-layered materials~\citep{wurdack2020two}, and large-area heterostructures~\citep{RN675, RN869, yang2015dual} more suitable for practical applications.

\subsection{Conclusion}
To conclude, recent theoretical and experimental research on proximity-coupled magnetic material--topological insulator heterostructures has demonstrated unexpectedly robust, intense magnetic proximity effects, and enabled the observation of exotic emergent phenomena like QAHE, axion insulator states, and skyrmions in topological systems. These achievements prompt new challenges: to engineer toplogical insulator--magnetic material heterostructures with higher temperature operation, and to demonstrate the fundamental operating principles underlying a new generation of topological devices. If these challenges can be overcome, then proximity-coupled topological insulator--magnetic material heterostructures can provide a powerful  platform for new low-energy spintronics and other technologies.

\begin{acknowledgements}
The authors acknowledge support by the Australian Research Council through the ARC Centre of Excellence in Future Low Energy Electronics Technologies (FLEET). M.G. acknowledges an Australian Government Research Training Program Scholarship. The authors would like to thank Prof. Dimitrie Culcer for the helpful discussions regarding QAHI edge states, and Mr. Alexander Nguyen for his help with proof-reading the paper.
\end{acknowledgements}

\bibliographystyle{MSP}


\end{document}